\documentclass[twocolumn]{aastex631}
\usepackage{amsmath}
\usepackage{graphicx}
\usepackage{subfigure}
\usepackage{booktabs}
\usepackage{tablefootnote}

\usepackage{xcolor}
\usepackage{hyperref}

\received{April 18, 2025}
\revised{August 31, 2025}
\accepted{September 2, 2025}

\shorttitle{Stellar Collision}

\shortauthors{Yu and Lai}

\graphicspath{{./}{}}

\begin{document}
\title{Binary Stars Approaching Supermassive Black Holes: Hydrodynamics of Stellar Collisions, Mass Fallback and Partial TDEs}

\correspondingauthor{Fangyuan Yu, Dong Lai}
\email{yufangyuan@sjtu.edu.cn, dl57@cornell.edu}

\author[0009-0004-6973-3955]{Fangyuan Yu}
\affiliation{Zhiyuan College, Shanghai Jiao Tong University, Shanghai 200240, People’s Republic of China}
\affiliation{Tsung-Dao Lee Institute, Shanghai Jiao Tong University, Shanghai 201210, People’s Republic of China}

\author[0000-0002-1934-6250]{Dong Lai}
\affiliation{Tsung-Dao Lee Institute, Shanghai Jiao Tong University, Shanghai 201210, People’s Republic of China}
\affiliation{Center for Astrophysics and Planetary Science, Department of Astronomy, Cornell University, Ithaca, NY 14853, USA}

\begin{abstract}
When binaries are injected into low-angular-momentum orbits around a central supermassive black hole (SMBH), various outcomes can occur, including binary tidal breakup, double stellar disruptions and stellar collision.
We use hydrodynamical simulations to study stellar collisions triggered by binary-SMBH encounters, examining both head-on and grazing collisions in deep ($\beta_b=5$) and gentle ($\beta_b=0.6$) encounters, where $\beta_b$ is the ratio of the binary tidal disruption radius to the binary pericenter distance to the SMBH.
Head-on collisions consistently result in appreciable mass loss ($\sim 5\%$) and a single merger remnant. 
Grazing collisions have varied outcomes. 
In gentle encounters, multiple collisions typically form a single remnant with minimal mass loss ($\lesssim 1 \%$). 
For deep encounters, the result depends on the specific collision parameters and stellar structure: $\gamma=5/3$ polytropic stars in our simulation produced two disturbed remnants, while solar-type stars (modeled with \texttt{MESA}) in our deep-grazing run formed a single merger remnant in a low-velocity collision.
All merger remnants feature extended envelopes, making them susceptible to partial tidal disruptions when they return to the SMBH.
The morphology and orbital energy distribution of collision-induced debris differ significantly from those of tidal disruption event (TDE) debris of single stars. 
Approximately half of the collision-generated debris falls back onto the SMBH, exhibiting a distinct time evolution of the fallback rate. 
We suggest that such mass loss and fallback can generate electromagnetic flares that mimic weak TDEs.

\end{abstract}


\section{Introduction} 
\label{sec:intro}

Almost all galaxies harbor a supermassive black hole (SMBH) at their center \citep[see][for a review]{KormendyHo2013ARAA}, and the surrounding nuclear clusters are expected to have high stellar densities. 
This unique environment leads to complex dynamical processes and diverse phenomena around the SMBH \citep[see][for a review]{Alexander2017ARAA}. 
A large fraction of stars in nuclear clusters are members of stellar binaries. 
The interactions (i.e., gravitational scatterings) between binaries and single stars tend to destroy relatively wide (or soft) binaries, leaving behind only tightly bound (or hard) binaries \citep[e.g.,][]{Heggie1975MNRAS,Hills1975AJ,Heggie2003}. 
These tight binaries can be scattered into low-angular-momentum orbits toward the SMBH by interactions with other stars. 
When this happens, the SMBH can tidally breakup the binary, resulting in one star becoming bound to the SMBH and the other being ejected as a hyper-velocity star (HVS) that can escape the galaxy \citep[e.g.,][]{Hills1988Nature,YuTremaine2003,Gould2003ApJ}. 
This process (``Hills mechanism") has been extensively studied as a significant source of HVSs and tightly bound stars in the Galaxy \cite[e.g.,][]{Gualandris2005MNRAS,Bromley2006ApJ,Ginsburg2006MNRAS,Sesana2007ApJ,Sari2009ApJ,Brown2018MNRAS}. 

When individual stars approach the SMBH on sufficiently tight orbits, they may be disrupted by tidal forces, leading to tidal disruption events (TDEs) \citep[e.g.,][]{Rees1988nature,Stone2020SSRv}. 
In recent years, stars bound to the SMBH following binary breakup have been linked to repeating partial TDEs \citep[e.g.,][]{cufari2022ApJL,Somalwar2023arXiv}. 
Such stars have also been associated with quasi-periodic eruptions (QPEs), which are periodic bursts of accretion activities that are thought to result from the interaction between bound stars and the SMBH or the accretion disk around the SMBH \citep[e.g.,][]{Lu2023MNRAS,Linial2023ApJ,Zhou2024aPRD,Zhou2024bPRD,Yao2025ApJ}. 

Taking the finite size of stars into account reveals additional phenomena beyond simple binary breakups. 
For example, double stellar tidal disruptions can occur when both components of a binary are torn apart by the SMBH’s tidal forces \citep[e.g.,][]{Mandel2015ApJL,Bradnick2017MNRAS,Bonnerot2019MNRAS}. 
Even when the stars are not tidally disrupted, stellar collisions may happen during binary-SMBH encounters when the stars are brought close to each other due to the gravitational influence of the SMBH \citep[e.g.,][]{Ginsburg2007MNRAS,Antonini2010ApJ,Antonini2011ApJ,Bradnick2017MNRAS,Paper1}. 
Such a collision results in stellar merger and mass loss, which may lead to accretion flares onto the SMBH. 
The stellar merger can also lead to the formation of exotic stars in nuclear star clusters.

In \citet{Paper1} (hereafter Paper I), we used three-body scattering experiments to characterize the key dynamics of binary tidal breakups, double stellar disruptions, and stellar collisions induced by binary-SMBH encounters. 
We analyzed the collision probabilities of binaries with different orbital properties and concluded that the collision probability can reach a few to 10’s percent as the binary's pericenter distance (relative to the SMBH) varies from outside the tidal sphere to within it.
Additionally, we found that the impact velocities of the colliding stars are comparable to the escape velocity of the star.

In this paper, we extend the work of Paper I by performing detailed hydrodynamical simulations of stellar collisions generated by binary-SMBH encounters.
Our goal is to address several questions that cannot be resolved by point-mass scattering experiments alone: 
What are the outcomes of such collisions under various conditions - do they result in the formation of a single merger remnant, or do they simply leave two stars highly perturbed? 
What is the post-collision structure of the stars? 
How much mass is lost during the collision, and could the fallback of this mass onto the SMBH produce a distinctive observable signature?

Our paper is organized as follows. 
In Section~\ref{sec:method and setup}, we describe the setup of our numerical simulations, including the initial conditions and methodology. 
In Sections~\ref{sec:results_poly} and~\ref{sec:results_MESA}, we present the results from our representative simulations  for a $\gamma=5/3$ polytrope (representing low-mass main-sequence stars with $M \lesssim 0.4 M_\odot$) and for a solar-like main-sequence (MS) star respectively. 
Finally, in Section~\ref{sec:summary and discussion}, we summarize our findings and comment on the results of Paper I based on our hydrodynamical simulations.

\section{Method and Setup}
\label{sec:method and setup}

We performed simulations of stellar collisions during binary-SMBH encounters using the smoothed particle hydrodynamics (SPH) code \textsc{Phantom} \citep{Phantom}.

Following Paper I, we adopt three key dimensionless ratios to characterize the encounters, defined as follows:
\begin{enumerate}
    \item $\beta_b \equiv r_{\rm tide}^b / r_p$, where $r_p$ is the pericenter distance of the binary relative to the SMBH, and
    \begin{equation}
        r_{\rm tide}^b \equiv a_b \Bigl( \frac{M_{\rm BH}}{m_{12}} \Bigr)^{1/3}
    \label{eq:rtideb}
    \end{equation}
    is the tidal radius of the binary. 
    Here $a_b$ is the binary's semi-major axis, $M_{\rm BH}$ is the SMBH mass, and $m_{12}$ is the total mass of the binary system.
    
    \item $\beta_\star \equiv r_{\rm tide}^\star / r_p$, where $r_{\rm tide}^\star$ is the tidal radius of an individual star in the binary, defined as 
    \begin{equation}
        r_{\rm tide}^\star \equiv R_\star \Bigl( \frac{M_{\rm BH}}{m_\star} \Bigr)^{1/3}.
        \label{eq:rtides}
    \end{equation}
    Here $R_\star$ and $m_\star$ are the radius and mass of the star, respectively.
    
    \item $\alpha \equiv a_b / R_{\rm col}$, where $R_{\rm col}$ is the separation between the two stars at the moment of collision.
    For the equal-mass binaries ($m_1=m_2$) considered in this paper, $R_{\rm col}=2R_\star$.
\end{enumerate}

\subsection{Simulation Setup}
\label{sec:setup}

Throughout this paper, we consider circular, equal-mass binary. In the first set of simulations (Runs 1–4 in Table~\ref{tab:simulations}), each star is modeled as a $\gamma = 5/3$ polytrope with a mass $m_\star = 1\,M_\odot$ and a radius $R_\star = 1\,R_\odot$. 
In the second set (Runs 5 and 6), we adopt more realistic stellar models corresponding to Middle-Age Main-Sequence (MAMS) stars with a mass of $1\,M_\odot$ and an age of $\approx 5 \rm Gyr$ and a radius of $R_\star\approx 1.016 R_\odot$. The model was evolved with an initial solar metallicity of $Z = 0.02$ using the one-dimensional implicit stellar evolution code Modules for Experiments in Stellar Astrophysics (\texttt{MESA}), version \texttt{r24.08.1} \citep{Paxton2011,Paxton2013,Paxton2015,Paxton2018,Paxton2019,Jermyn2023}. For reproducibility, we also report the central properties of the star: the central density is $\rho_c\approx162.8 \rm{g\, cm^{-3}}$, and the central mass fractions of hydrogen and helium are $X_C\approx0.309$ and $Y_C\approx0.670$, respectively.
The binaries in our simulations are then created by shifting the centers of two stellar models.
For the subsequent hydrodynamical evolution, we adopt the gamma equation of state (EOS) with  $P = (\Gamma - 1)\rho u$  , where $P$, $\rho$, and $u$ are the pressure, density, and specific internal energy, respectively, and we set the index $\Gamma = 5/3$. 
This EOS is well-suited for capturing the essential dynamics of the collision, including shock heating.
Each model is relaxed  in isolation so that it reaches hydrostatic equilibrium  (We fix the radial entropy profile of the star during the relaxation. We stop the relaxation procedure when the ratio of kinetic to potential energy drops below $10^{-7}$ and the error in density is less than 1\%). 
In each simulation, the binary is placed on a Newtonian parabolic orbit around a $10^6 M_\odot$ SMBH from a distance of $(2-3)\, r_{\rm tide}^b$. 
Note that the probability of stellar collision depends on the orbital phase of the binary.
To ensure collision, we adjust the initial binary orbital phase by evolving the binary for certain duration before releasing it on the orbit toward the SMBH.

For the penetrating factor $\beta_b$, we select $\beta_b = 0.6$ to represent a gentle encounter ($\beta_b < 1$) and $\beta_b = 5$ for a deep encounter ($\beta_b > 1$). 
Throughout the paper, we consider initially coplanar, circular binaries with $\alpha = 8$ , which corresponds to $a_b=16R_\odot$.
(see Paper I for the effect of binary eccentricity on the collision probability).

We have performed simulations with $2 \times 10^5$, $10^6$, and $2 \times 10^6$ SPH particles, and found very little deviations in the simulation results (see Figures~\ref{fig:masslossfrac},\ref{fig:fallbackrate}).
Unless otherwise noted, all results presented in this paper are from the $10^6$ particle runs.

We run each simulation for a few hundred stellar dynamical time ($t_{\rm dyn}\equiv \sqrt{R_\star^3/Gm_\star}$) after the collision to ensure that the energies of the collision remnants have converged and the final collision outcomes are clearly determined.

\begin{table*}[htbp]
    \centering
    \begin{tabular}{cccccccc}
      \toprule
        Runs & $\beta_b$ & $N$ & $r_{12} (R_\odot)$ & $|v_{12}| (\sqrt{GM_\odot/R_\odot})$ & ${b_{12}/r_{12}}$ & Number of remnant(s) & ${\Delta M / M_{\rm tot}}$\\ 
      \midrule
        1a & 5 & $5\times10^5$ & $3.749$ & $1.155$ & $0.0104$ & $1$ & $0.0495$ \\
        1b & 5 & $1\times10^5$ & $3.810$ & $1.171$ & $0.0625$ & $1$ & $0.0538$ \\
        1c & 5 & $1\times10^6$ & $3.723$ & $1.147$ & $0.0484$ & $1$ & $0.0492$ \\
        2 & 5 & $5\times10^5$ & $3.612$ & $1.354$ & $0.4885$ & $2$ & $ 0.0375$ \\
        3 & 0.6 & $5\times10^5$ & $3.329$ & $1.046$ & $0.0285$ & $1$ & $0.0476$ \\
        4 & 0.6 & $5\times10^5$ & $2.388$ & $1.200$ & $0.9085$ & $1$ & $ 0.0068$ \\ 
     \midrule
         5 &  5 &  $5\times10^5$ &  $4.074$ &  $1.186$ &  $0.1469$ &  $1$ &  $0.0720$ \\
         6 &  5 &  $5\times10^5$ &  $3.855$ &  $1.012$ &  $0.525$ &  $1$ &  $0.0159$ \\
      \bottomrule
    \end{tabular}
    
    \caption{Summary of the SPH simulations.
    Here $\beta_b=r_{\rm tide}^b/r_p$ (with $r_{\rm tide}^b$ the binary tidal radius given by Eq.~(\ref{eq:rtideb}) and $r_p$ the pericenter distance), and $N$ refers to the number of SPH particles used to represent each star,
    $r_{12}$ and $v_{12}$ are the relative separation and velocity just before collision$^\dagger$,
    and $b_{12} \equiv |\vec{v}_{12} \times \vec{r}_{12}| / |\vec{v}_{12}|$ is the impact parameter. 
    We also list the number of remnant(s) and the mass loss fraction $\Delta M / M_{\rm tot}$ (where $M_{\rm tot}=2m_\star$) at the end of the simulation. 
    The mass loss fractions are all calculated using the  iterative Bernoulli method (see Section~\ref{sec:identify}  below).
    In the first set of simulations (Runs 1–4), each star is modeled as a $\gamma = 5/3$ polytrope. In the second set (Runs 5, 6), each star is modeled as a MAMS star computed from \texttt{MESA}.
    For the deep, grazing cases, the contact conditions are not identical across models (see Runs 2,6). These differences plausibly contribute to the distinct outcomes alongside stellar structure.}
    \label{tab:simulations}
    \vspace{0em}
    \begin{minipage}{\linewidth}
        \footnotesize
        $^\dagger$ Note that it is not possible to determine the exact moment of collision. In our analysis, we use the first numerical output after the collision occurs as the approximate collision time ($t = 0$), and $r_{12}$ and $v_{12}$ are measured at $t\approx-2t_{\rm dyn}$ (with $t_{\rm dyn}\equiv\sqrt{R_\star^3/Gm_\star}$).
    \end{minipage}
\end{table*}

Our representative simulations are summarized in Table~\ref{tab:simulations}.

\subsection{Identify Post-collision Outcomes}
\label{sec:identify}

In Paper I, we adopted the simplest assumption for the collision process, treating it as a perfectly inelastic collision when the separation between the two stars becomes smaller than $R_{\rm col}$, resulting in a single collision remnant. 
However, in our SPH simulations, we observe that collisions can lead to some degree of mass loss, and it is also possible that the two stellar remnants do not merge. 
From the wide range of simulations we have carried out, we select four for detailed presentation in Section~\ref{sec:results_poly} (see Table~\ref{tab:simulations}).
 These represent four general scenarios that characterize the diversity of outcomes in our parameter space.

\begin{enumerate}
    \item {\it Head-on collision in a deep encounter ($\beta_b=5$).}
    The stars collide when the binary reaches the pericenter to the SMBH. 
    If the collision occurs with low relative velocity and a small impact parameter, the stars may merge, forming a new, more massive star.

    \item {\it Grazing collision in a deep encounter ($\beta_b=5$).}
    In a deep encounter, the binary is disrupted after the pericenter passage. 
     High-velocity, large-impact-parameter encounters that avoid stellar merger typically lead to an outcome that is similar to the Hills mechanism of binary breakup: one star is ejected, while the other remains bound to the SMBH. 
    A key difference is that both interacting stars may suffer significant structural disturbances during the grazing collision. 
    In contrast, low-velocity encounters can result in a merger even at large impact parameters. 
    In this work, whether a merger occurs depends on the specific values of $v_{12}$ and $b_{12}/r_{12}$ at contact (see Table~\ref{tab:simulations}), rather than the stellar structure alone.
    Such a merger forms a new, more massive star.

    \item {\it Head-on collision in a gentle encounter ($\beta_b=0.6$).} 
    In a gentle encounter, the binary survives the pericenter passage. 
    However, the binary can gain a sufficiently large eccentricity, which leads to stellar collisions after the pericenter passage. 
    If the two stars collide with low relative velocity and small impact parameter (a head-on collision), they may coalesce to form a new, more massive star.

    \item {\it Grazing collision in a gentle encounter ($\beta_b=0.6$).}
    In a grazing collision, where the impact parameter is relatively large, the stars may not merge but instead experience significant perturbations, such as mass exchange or mass loss. 
    In a gentle encounter, the binary survives the pericenter passage, allowing for repeated collisions. 
    This increases the likelihood of a merger, even if the first collision does not result in an immediate merger.

\end{enumerate}

A qualitative description of these simulation outcomes (e.g., whether a collision occurs or how many remnants exist) is relatively straightforward. 
However, when performing a quantitative analysis (e.g., to determine the mass loss), careful consideration is required to distinguish SPH particles belonging to the remnant and those contributing to mass loss, particularly due to the exchange of particles between the stars during the collision.

We employ an iterative scheme to quantitatively classify SPH particles as belonging to one or two gravitationally bound stellar cores ($S_k$) or as part of the unbound mass loss. 
The algorithm begins with an initial assignment of particles to potential cores and then iteratively refines this classification until convergence is achieved. The following describes each step of this procedure in detail.

To initiate the iterative process at a given snapshot, we first require an initial assignment of particles to potential cores. The procedure differs for single-core and dual-core cases:

\begin{itemize}
    \item \textbf{Single-core case}:
    \begin{enumerate}
        \item Compute the global center-of-mass, $\vec{r}_{\rm CM,global}$, using all SPH particles.
        \item Assign all particles $i$ with $|\vec{r}_i - \vec{r}_{\rm CM,global}| \le R_{\rm guess}$ as the initial members of the single core $S_1$.
    \end{enumerate}

    \item \textbf{Dual-core case}:
    \begin{enumerate}
        \item Construct a 2D grid of the projected column density (since the orbits are coplanar, the center-of-mass lies in the $z=0$ plane). Ensure the grid resolution is finer than $0.1\,R_\star$.
        \item Identify the two density peaks from the 2D grid as the initial core centers, denoted by $\vec{r}_{\rm CM,init}^{(1)}$ and $\vec{r}_{\rm CM,init}^{(2)}$.
        \item For each SPH particle $i$, calculate its distances to the two centers: $d_i^{(1)} = |\vec{r}_i - \vec{r}_{\rm CM,init}^{(1)}|$ and $d_i^{(2)} = |\vec{r}_i - \vec{r}_{\rm CM,init}^{(2)}|$.
        \item Assign particle $i$ to core $S_1$ if $d_i^{(1)} \le R_{\rm guess}$ and either $d_i^{(2)} > R_{\rm guess}$ or $d_i^{(1)} \le d_i^{(2)}$.
        \item Otherwise, assign particle $i$ to core $S_2$ if $d_i^{(2)} \le R_{\rm guess}$ and either $d_i^{(1)} > R_{\rm guess}$ or $d_i^{(2)} < d_i^{(1)}$.
    \end{enumerate}
\end{itemize}

The parameter $R_{\rm guess}$ can be adjusted to optimize the initial identification performance; we adopt $R_{\rm guess}=1 R_\star$ for single-core cases and $R_{\rm guess}=2 R_\star$ for dual-core cases.

Once the cores are initialized, the main iterative loop begins. In each iteration, we first update the core properties.
For every identified core $k$ (with $k=1$ for the single-core case, and $k=1,2$ for the dual-core case), we first compute the following quantities based on the current set of constituent particles $S_k$:

\begin{itemize}
    \item \textbf{Number of particles ($N_k$)}: The number of SPH particles currently assigned to core $S_k$. Given equal particle mass $m_p$, the total mass of the remnant is $M_k = N_k \times m_p$.
    \item \textbf{Center-of-mass position $\vec{r}_{\rm CM}^{(k)}$ and velocity $\vec{v}_{\rm CM}^{(k)}$} of the core:
    \begin{equation}
        \vec{r}_{\rm CM}^{(k)} = \frac{1}{N_k} \sum_{i \in S_k} \vec{r}_i,\quad \vec{v}_{\rm CM}^{(k)} = \frac{1}{N_k} \sum_{i \in S_k} \vec{v}_i.
    \end{equation}
\end{itemize}

Our method for identifying the post-encounter remnant is based on the Bernoulli constant.
For each SPH particle $i$ belonging to the remnant $k$, a specific enthalpy is calculated according to:
\begin{equation}
    h_i^{(k)}=\frac{1}{2}(\vec{v}_i-\vec{v}^{(k)}_{\rm CM})^2+u_i+\frac{P_i}{\rho_i}+\phi_i^{(k)},
    \label{eq:enthalpy}
\end{equation}
where $u_i$ and $\rho_i$ are its specific internal energy and density, $P_i$ is the pressure, and $\phi_i^{(k)}$ is the gravitational potential due to the mass distribution of core $k$ excluding the influence of SMBH.
Along a streamline with no dissipation, $h_i^{(k)}$ should be a constant according to the Bernoulli theorem.
Particles with $h_i^{(k)}>0$ do not belong to $S_k$ - they are considered as the mass loss, unbound from the remnant.

The kinetic and internal energy terms in Equation (\ref{eq:enthalpy}) are readily available from the simulation output. However, the evaluation of $\phi_i^{(k)}$ is more challenging. While the most accurate approach would be a direct summation of the potential contributions from all particles $j \in S_k$, this method is computationally prohibitive for large $N$. Therefore, we adopt two approximate methods for estimating $\phi_i^{(k)}$, tailored for the single-core and dual-core cases:

\begin{itemize}
    \item \textbf{Global-approximation for the single-core case}:\\
    This approach estimates the gravitational potential from the remnant by subtracting the known SMBH contribution $\phi_{\text{BH},i}$ from the total potential at particle $i$:
	\begin{equation}
	    \phi_i^{(k)} \simeq \phi_{\text{total},i} - \phi_{\text{BH},i},
	\end{equation}
    where $\phi_{{\rm total},i}$ is provided by the SPH simulation output.

    \item \textbf{Point-mass-approximation for the dual-core case}:\\
    Here, core $k$ is modeled as a point mass $M_k = N_k m_p$ located at its center-of-mass $\vec{r}_{\rm CM}^{(k)}$. Thus,
	\begin{equation}
	    \phi_i^{(k)} \approx -G \frac{M_k}{|\vec{r}_i - \vec{r}_{CM}^{(k)}|}.
	\end{equation}
    To avoid singularities, a softening length or a lower bound on $|\vec{r}_i - \vec{r}_{\rm CM}^{(k)}|$ (e.g., $>\epsilon$) is applied.
\end{itemize}

After calculating the specific enthalpy $h_i^{(k)}$, we re-assign each particle according to the following criteria:

\begin{itemize}
    \item \textbf{Single-core case:} If $h_i^{(1)} < 0$, particle $i$ is considered gravitationally bound and is assigned to (or retained in) core $S_1$. Otherwise, it is classified as unbound.
    
    \item \textbf{Dual-core case:} Particle $i$ is assigned to core $S_k$ (with $k=1$ or $2$) for which $h_i^{(k)}$ is minimized, provided that the minimum $h_i^{(k)} < 0$. If $h_i^{(k)} \ge 0$ for all cores, the particle is considered unbound.
\end{itemize}

The sets of particles $\{S_k\}$ are then updated based on the new assignments, and the process is repeated until convergence is achieved.

Although most of our calculations converge within a few iterations, we implement an additional optimization to further accelerate convergence during the later stages, when the cores are already well established and changes in membership occur primarily at the outskirts of the remnant.
After a specified number of initial full iterations, particles currently assigned to core $k$ and located within a small radius $R_{\rm frozen}$ (e.g., $R_{\rm frozen} = R_\star$) of that core's center-of-mass $\vec{r}_{\rm CM}^{(k)}$ are designated as \textit{frozen}. 
These particles are assumed to remain bound and are excluded from subsequent enthalpy evaluations and re-assignment steps. 
In each following iteration, only particles outside the frozen region or those previously classified as unbound are re-evaluated. 
Nevertheless, the center-of-mass position and total mass (or particle count) of each core are still computed using the complete set of its current members, including both frozen and re-assigned particles. 
This strategy significantly reduces the number of particles requiring costly enthalpy calculations at each step, thereby improving computational efficiency without sacrificing accuracy.

Once the iterative process converges, any particle not assigned to a bound core ($S_k$) is classified as mass loss. 
For subsequent analysis, we record the indices of all SPH particles, distinguishing between those assigned to bound cores and those identified as unbound.

\subsection{Calculation of Mass Fallback Rate}
\label{sec:CalculationMdot}
Once we have determined which SPH particles are mass loss, we can calculate the time it takes for them to fall back to SMBH (to the pericenter) to determine the mass fallback rate $\dot{M}$.

An approximate method of calculating $\dot{M}$ is as follows.
Our hydrodynamical simulations allow us to calculate the  specific binding energy $\varepsilon$ of the mass-loss particles relative to the SMBH and to determine their energy distribution $dM/d\varepsilon$.
This distribution can be used to determine the mass fallback rate $\dot{M}$ as a function of time through Kepler’s third law:
\begin{equation}
    \dot{M}(t) = \frac{dM}{d\varepsilon} \frac{d\varepsilon}{dt} = \frac{(2\pi GM_{\rm BH})^{2/3}}{3}  \frac{dM}{d\varepsilon} t^{-5/3}.
\label{eq:tdefallback}
\end{equation}
This approach is highly effective for calculating the mass fallback rate in TDEs, as the star is disrupted at pericenter, resulting in a narrow angular momentum distribution. 
Consequently, nearly all bound particles must complete a full orbit before returning to pericenter.

However, the above approach for calculating $\dot{M}$ does not always work well for stellar collisions considered here. 
In deep encounters, collisions typically occur near the pericenter, but in gentle encounters, collisions tend to happen farther away, making Eq.~(\ref{eq:tdefallback}) less accurate. 
Additionally, the mass loss caused by stellar collisions can have a broad angular momentum distribution, implying that the particles do not necessarily need to complete an entire orbit to return to the pericenter.
Thus, it is important to analyze the trajectory of each particle individually to accurately determine the mass fallback rate $\dot{M}$.

To compute the pericenter distance $r_p$ and fallback time $t_{\text{fb}}$ for each SPH particle, we analyze the orbital parameters derived from the position $\mathbf{r}$ and velocity $\mathbf{v}$ of the particle.
The specific angular momentum vector is $\mathbf{h} = \mathbf{r} \times \mathbf{v}$ and the specific orbital energy (relative to the SMBH) is
\begin{equation}
\varepsilon = \frac{v^2}{2} - \frac{GM_{\rm BH}}{r}.
\end{equation}
 
The semi-major axis, orbital eccentricity and pericenter distance are then given by
\begin{equation}
\quad a = -\frac{GM_{\rm BH}}{2\varepsilon}, e = \sqrt{1 - \frac{h^2}{GM_{\rm BH}a}}, \quad r_p = \frac{h^2}{GM_{\rm BH}(1+e)}.
\end{equation}
To determine the angle $\theta$ between the particle’s instantaneous position and its pericenter direction, we compute the eccentricity vector
\begin{equation}
\mathbf{e} = \frac{\mathbf{v} \times \mathbf{h}}{GM_{\rm BH}} - \frac{\mathbf{r}}{r},
\end{equation}
which points from the SMBH toward the orbital pericenter. The angle $\theta$ is then obtained via
\begin{equation}
\cos\theta = \frac{\mathbf{e} \cdot \mathbf{r}}{e r}.
\end{equation}

For bound orbits ($e < 1$), the eccentric anomaly ${\cal{E}}$ and the mean anomaly ${\cal{M}}$ are 
\begin{equation}
\cos {\cal{E}} = \frac{e + \cos\theta}{1 + e\cos\theta}, \quad {\cal{M}} = {\cal{E}} - e\sin {\cal{E}},
\end{equation}
and the fallback time is 
\begin{equation}
t_{\text{fb}} =  \sqrt{\frac{a^3}{GM_{\rm BH}}}(2\pi - {\cal{M}}).
\end{equation}

For hyperbolic orbits ($e > 1$), the hyperbolic anomaly ${\cal{H}}$ and the mean anomaly ${\cal{M}}$ are 
\begin{equation}
\cosh {\cal{H}} = \frac{e + \cos\theta}{1 + e\cos\theta}, \quad {\cal{M}} = e\sinh {\cal{H}} - {\cal{H}},
\end{equation}
with the fallback time given by (if it can reach pericenter)
\begin{equation}
t_{\text{fb}} = \sqrt{\frac{(-a)^3}{GM_{\rm BH}}} {\cal{M}}.
\end{equation}

In practice, we measure $\mathbf{r}$ and $\mathbf{v}$ of the unbound (relative to the remnant) SPH particle at a time after stellar collision when the mass loss fraction has stabilized. 
We then calculate the fallback times for all mass-loss particles that will return to the SMBH. 
Finally, we manually add the approximate time from the collision to the calculated fallback time and sum over the mass-loss particles to obtain the final mass fallback rate.

\section{\texorpdfstring{Results: $\gamma = 5/3$ polytrope}{Results: gamma = 5/3 polytrope}}
\label{sec:results_poly}

Here we present our numerical results for $\gamma=5/3$ polytrope stars, representing $M\lesssim 0.4M_\odot$ MS stars.

For Run 1, we record the results with different resolutions (i.e., the number of SPH particles) to test the consistency of the results. 
We also record the relative distance $|\vec{r}_{12}|$ and velocity $|\vec{v}_{12}|$ between the two stars approximately $2 t_{\rm dyn}$ (where $t_{\rm dyn}\equiv\sqrt{R_\star^3/Gm_\star}$) before the collision and compute the impact parameter, defined as $b_{12} \equiv |\vec{v}_{12} \times \vec{r}_{12}| / |\vec{v}_{12}|$,  to characterize the nature of the encounter. 
Throughout this paper, we define a collision as ``nearly head-on" if the impact parameter is a small fraction of the contact separation ($b_{12}/r_{12}\lesssim 0.15$), and as ``grazing" if the impact parameter is a significant fraction of it ($b_{12}/r_{12}\gtrsim 0.5$). 
For brevity, we will hereafter refer to ``nearly head-on" collisions simply as ``head-on" collisions.
Additionally, at the end of the simulation, we record the number of merger remnants and the mass loss fraction computed using the method described in Section~\ref{sec:identify}. 
Note that when two merger remnants are present, we use the mass radial profile centered on each remnant's centroid to estimate the amount of mass loss.

\begin{figure*}[htbp]
    \centering
    \begin{minipage}[b]{0.49\linewidth}
        \centering
        \includegraphics[width=1\textwidth]{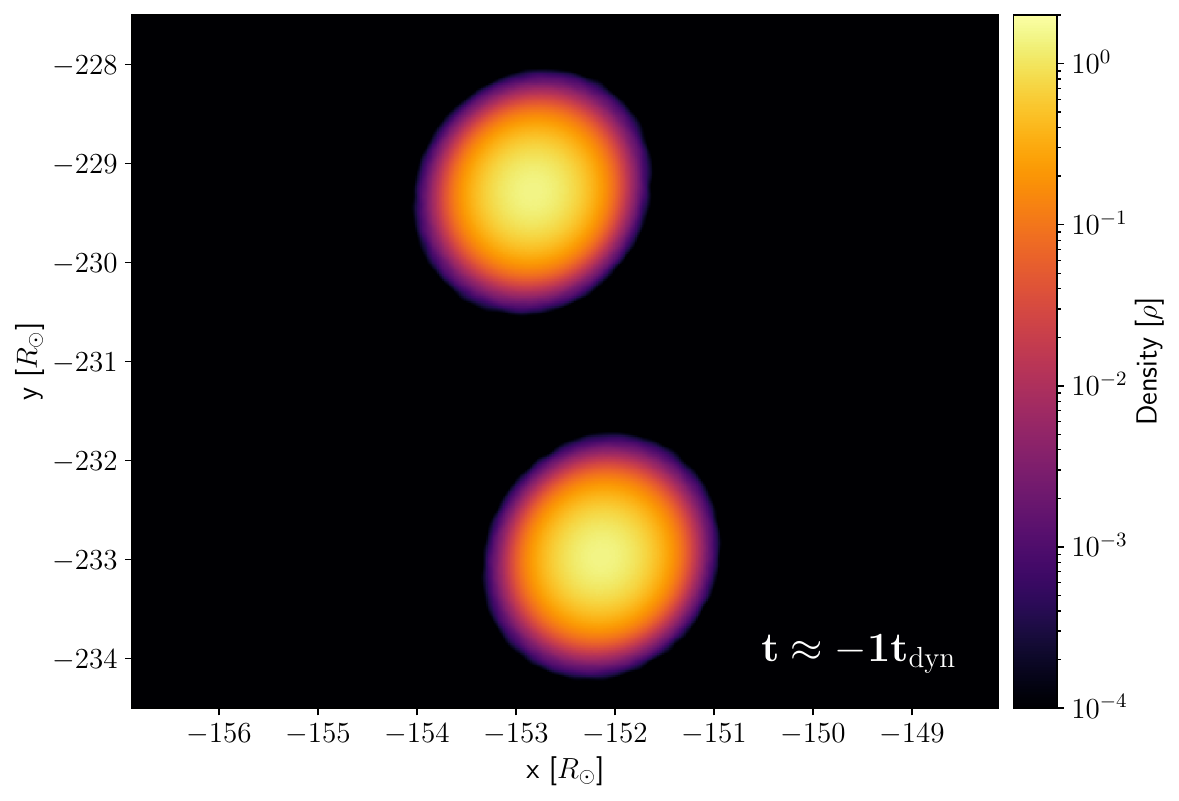}
    \end{minipage}
    \begin{minipage}[b]{0.49\linewidth}
        \centering
        \includegraphics[width=1\textwidth]{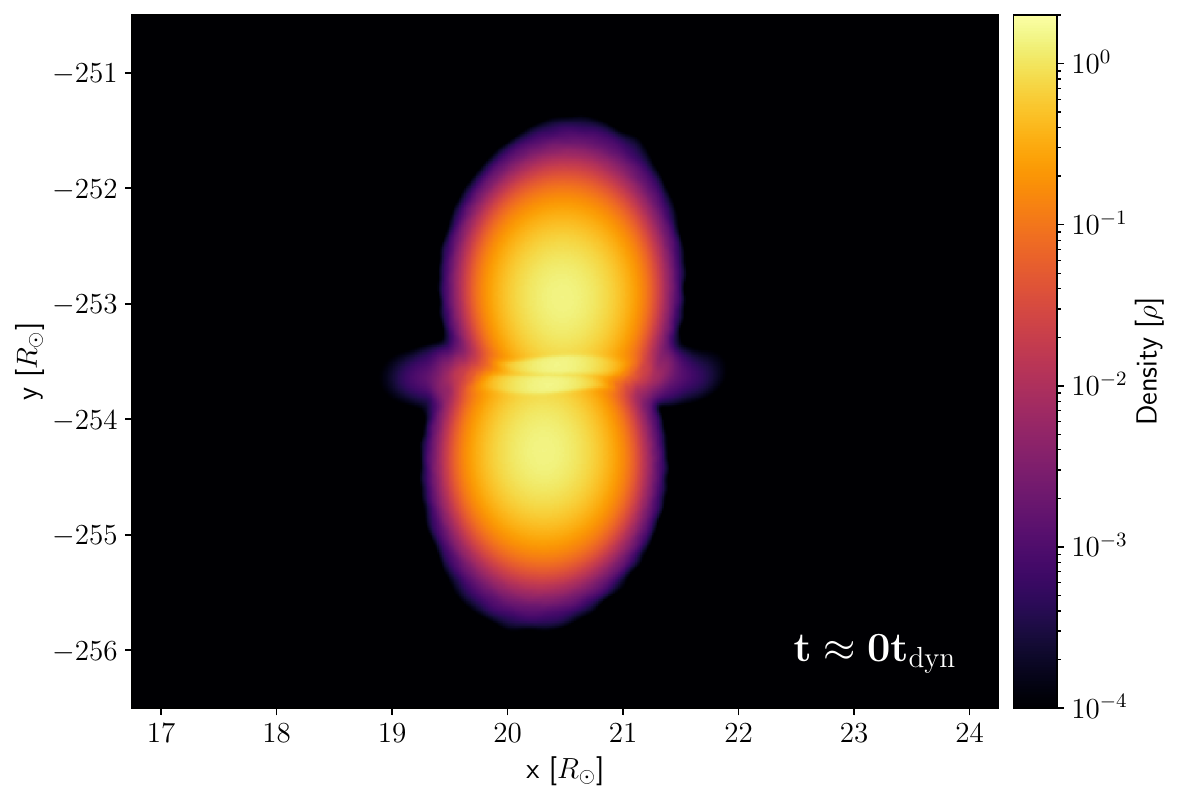}
    \end{minipage}
    \begin{minipage}[b]{0.49\linewidth}
        \centering
        \includegraphics[width=1\textwidth]{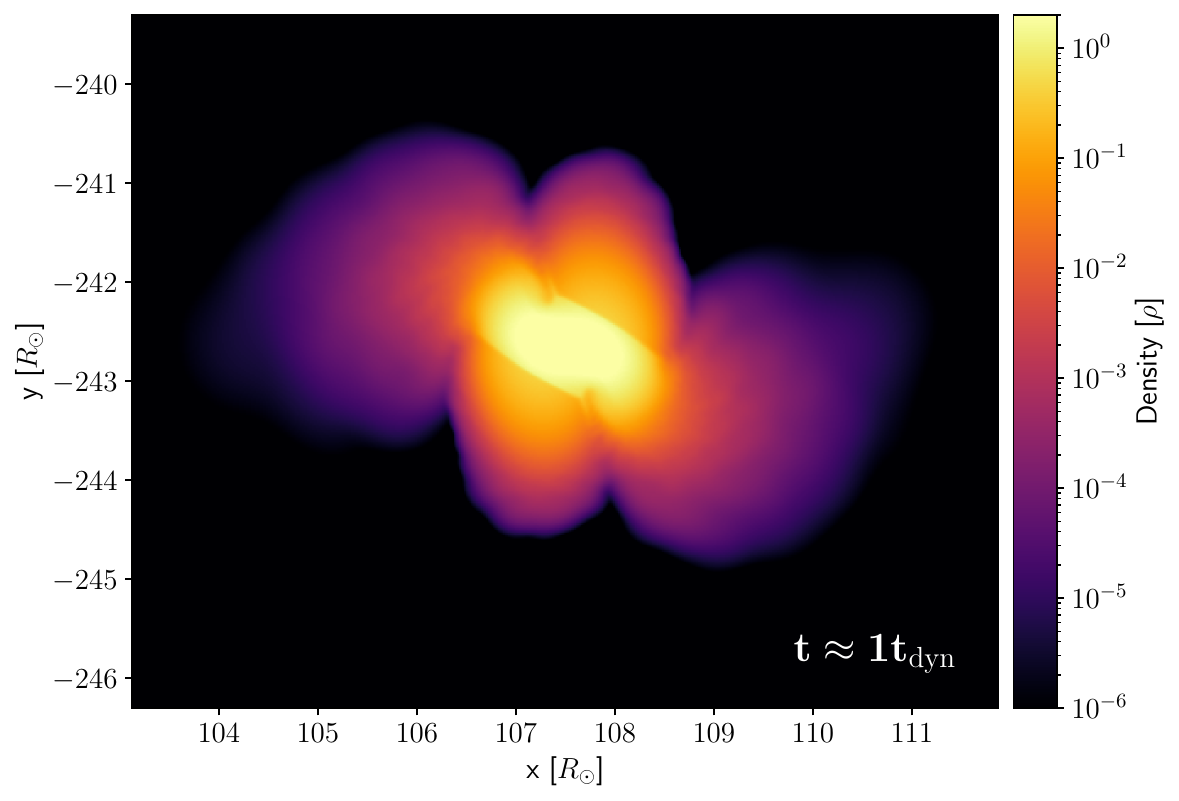}
    \end{minipage}
    \begin{minipage}[b]{0.49\linewidth}
        \centering
        \includegraphics[width=1\textwidth]{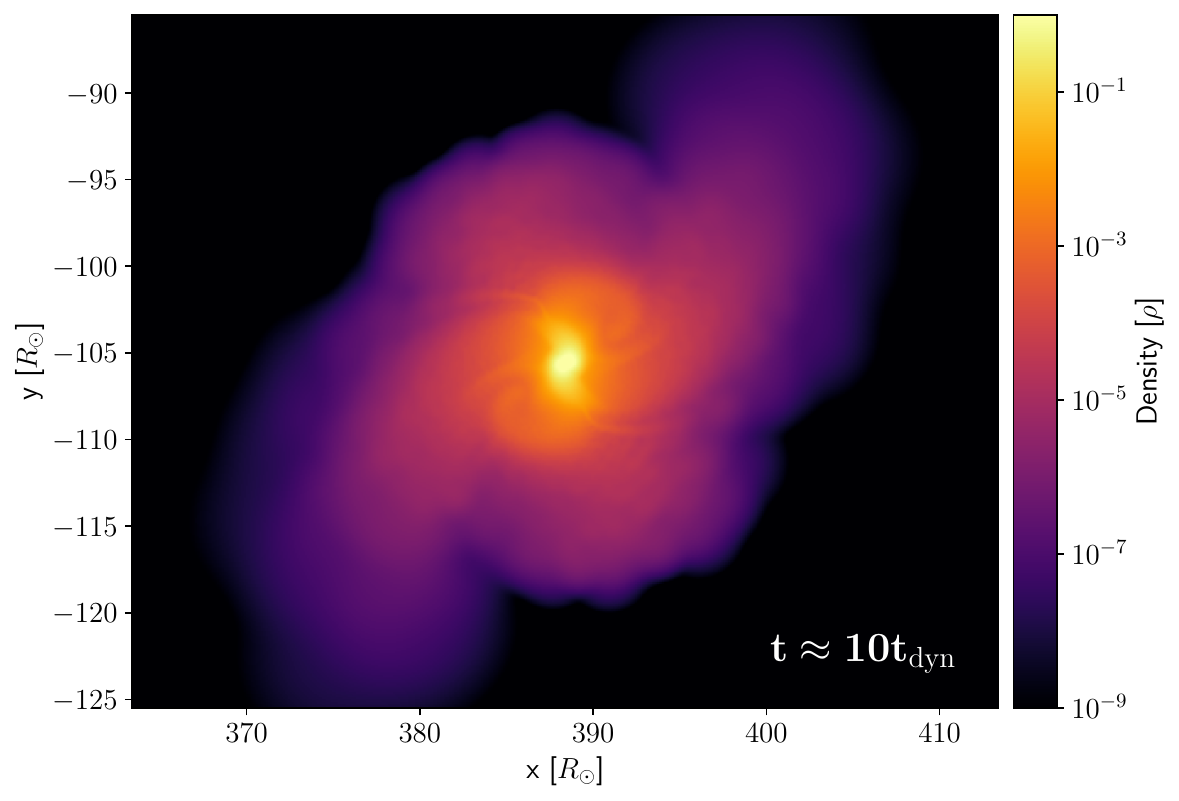}
    \end{minipage}

    \begin{minipage}[b]{0.49\linewidth}
        \centering
        \includegraphics[width=1\textwidth]{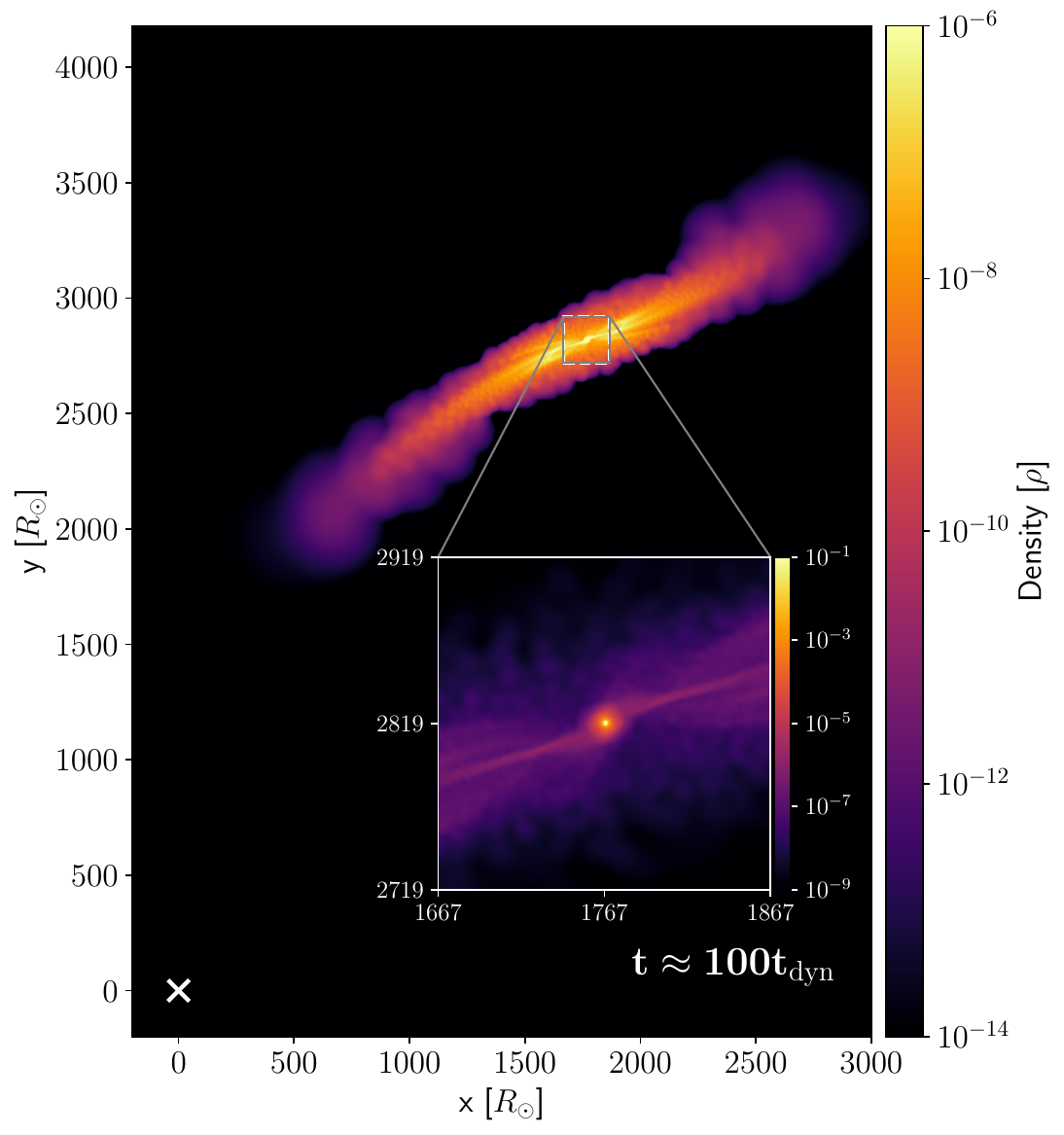}
    \end{minipage}
    \begin{minipage}[b]{0.49\linewidth}
        \centering
        \includegraphics[width=1\textwidth]{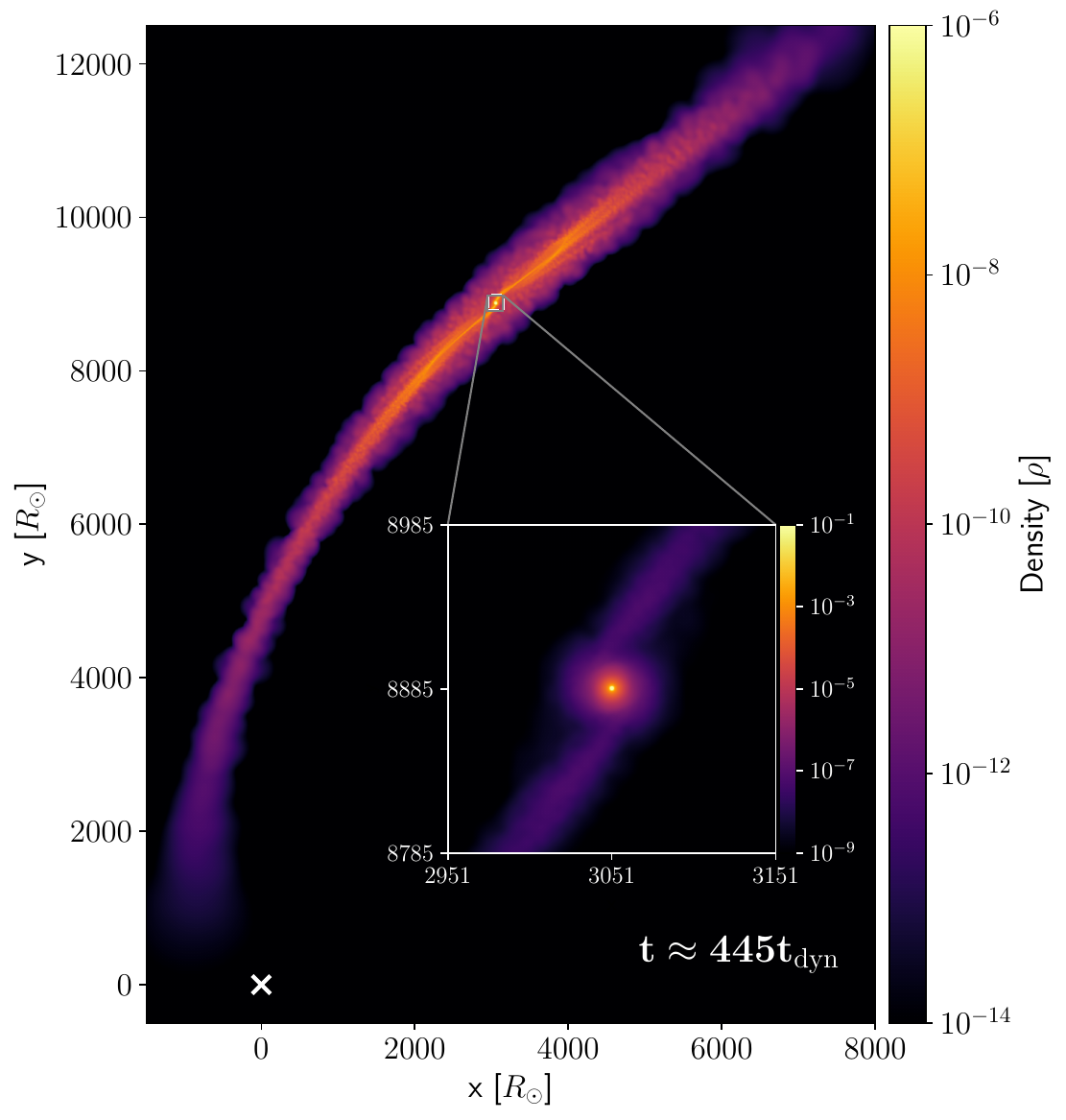}
    \end{minipage}
    \caption{The density distribution plots of Run 1a at different times for the $z=0$ slice (the orbital plane). 
    The density values in the plot are in visualization code units. 
    Note that the maximum value of the color bar does not represent the actual maximum value in the plot, but is instead chosen for better visualization of the debris stream's overall structure. 
    In the bottom row, the small inset square shows a zoomed-in version of the region inside the dashed box from the larger plot, with the color bar adjusted for better display of the central merger remnant. 
    The white ``$\times$'' marks the position of the SMBH. 
    Here, $t = 0$ approximately corresponds to the moment of collision.}
    \label{fig:deepheadonmovie}
\end{figure*}

\begin{figure*}[htbp]
    \centering
    \begin{minipage}[b]{0.49\linewidth}
        \centering
        \includegraphics[width=1\textwidth]{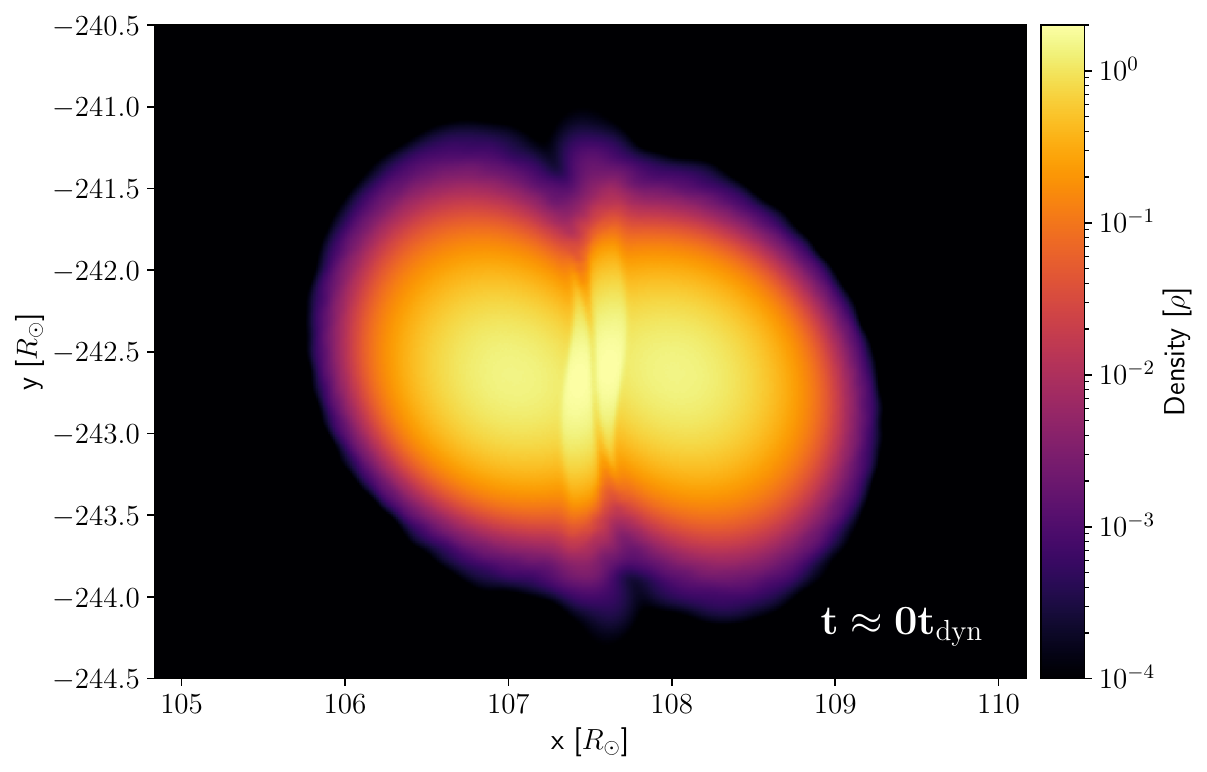}
    \end{minipage}
    \begin{minipage}[b]{0.49\linewidth}
        \centering
        \includegraphics[width=1\textwidth]{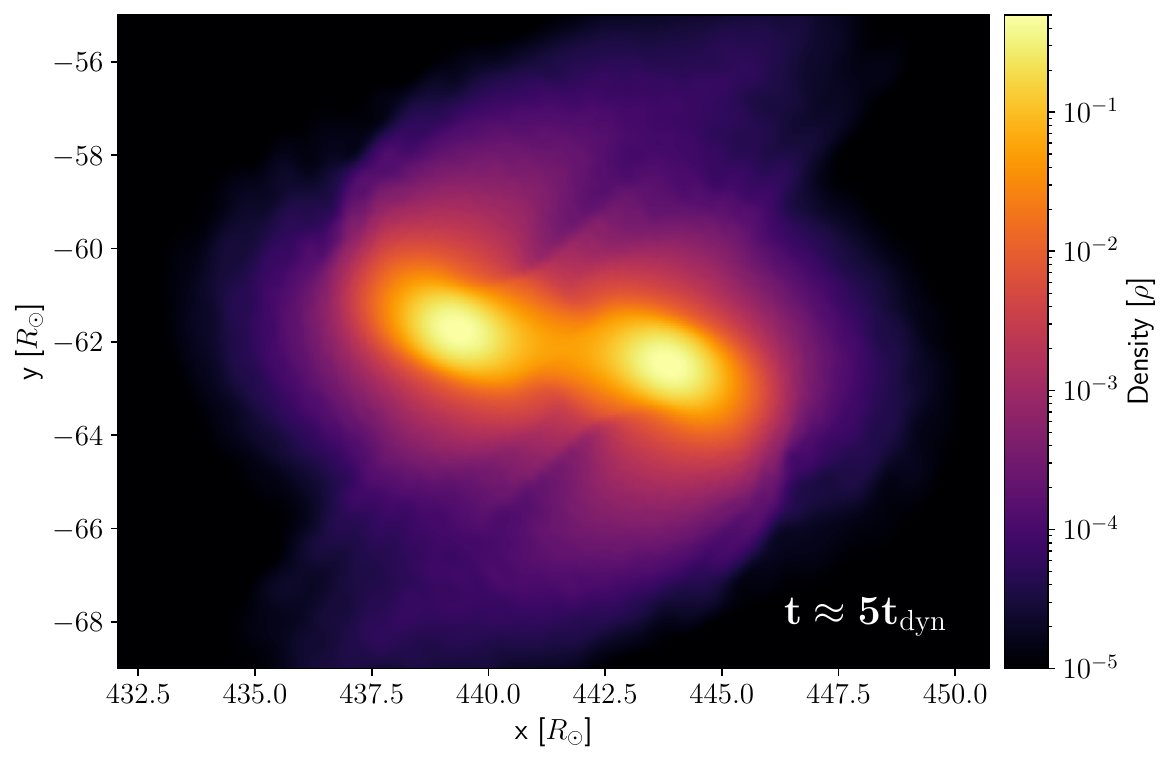}
    \end{minipage}
    \begin{minipage}[b]{0.49\linewidth}
        \centering
        \includegraphics[width=1\textwidth]{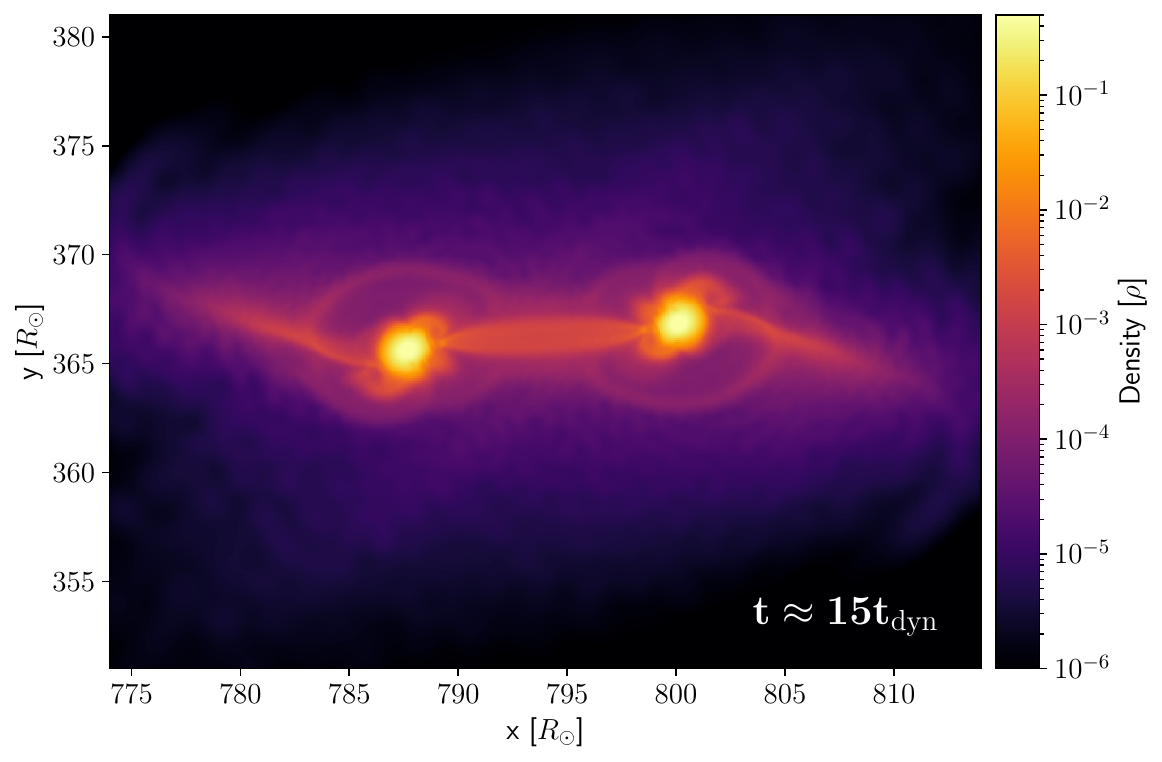}
    \end{minipage}
    \begin{minipage}[b]{0.49\linewidth}
        \centering
        \includegraphics[width=1\textwidth]{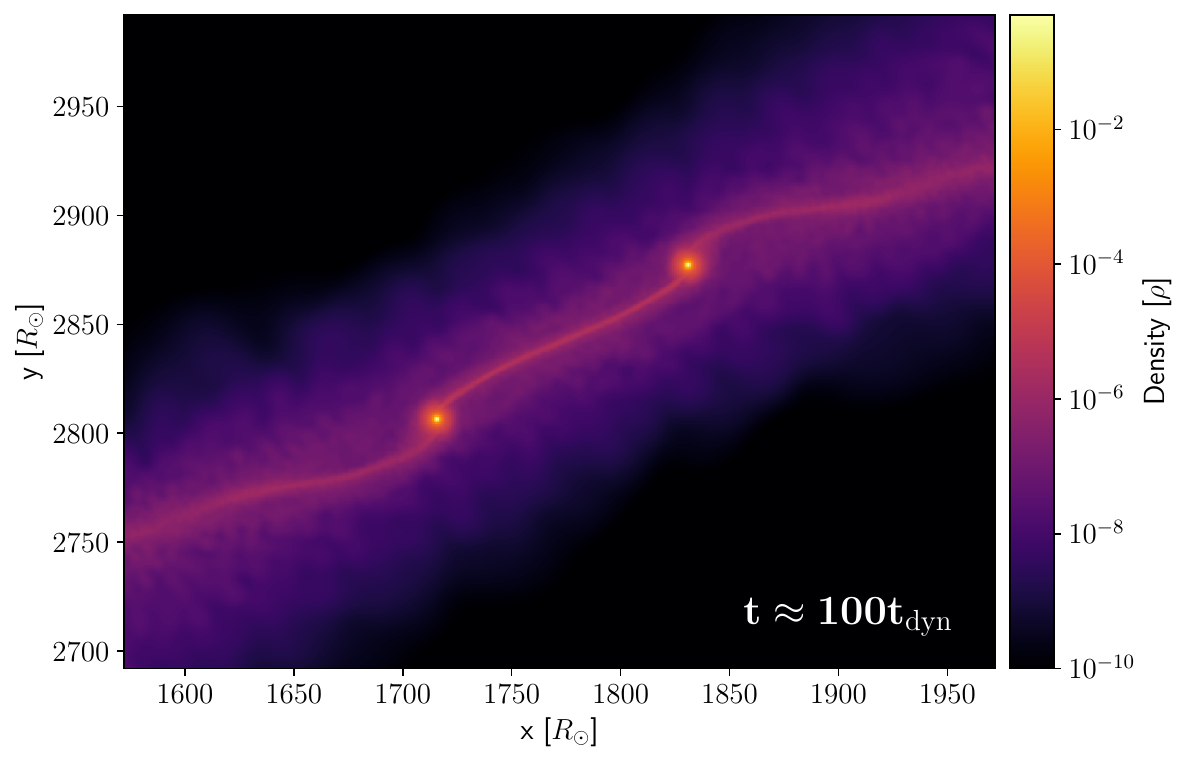}
    \end{minipage}

    \begin{minipage}[b]{0.49\linewidth}
        \centering
        \includegraphics[width=1\textwidth]{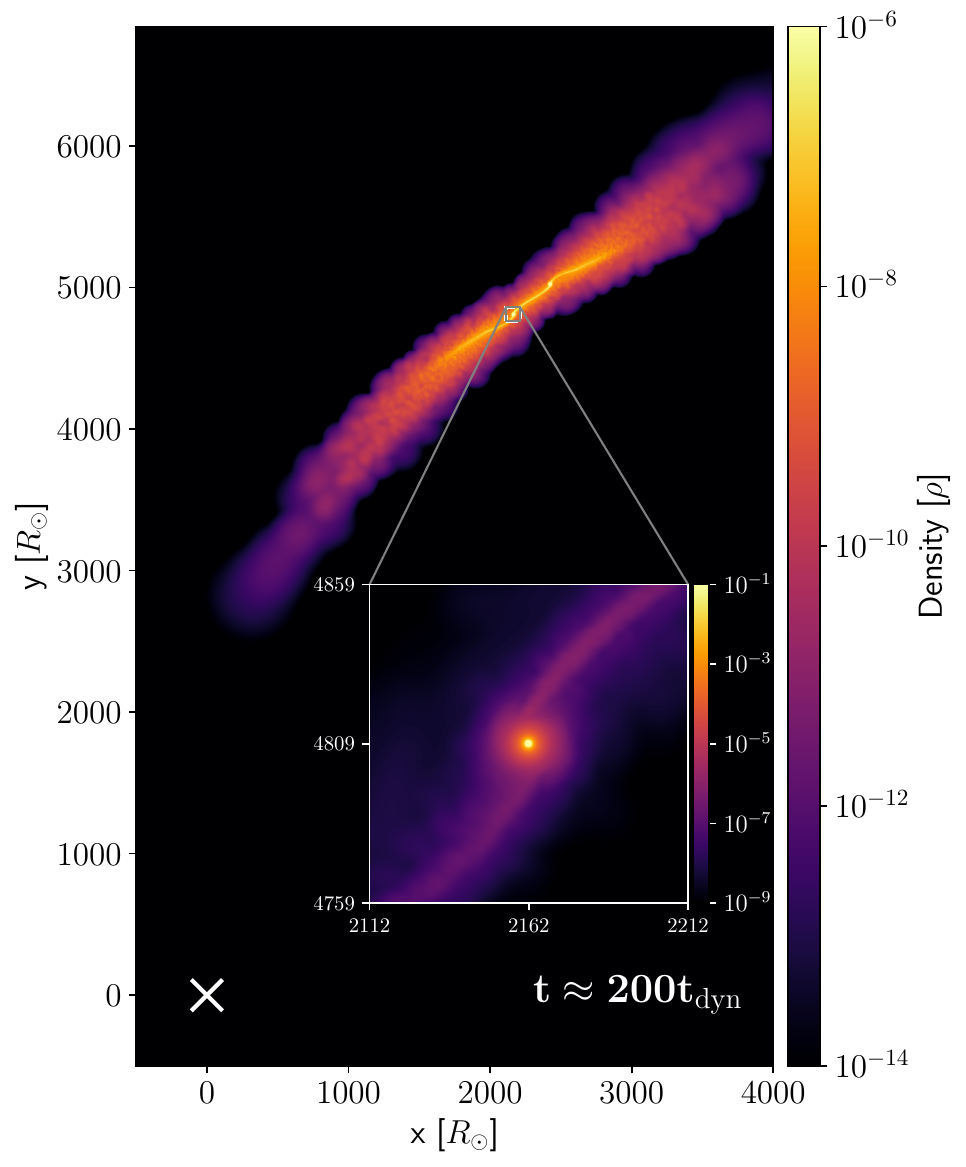}
    \end{minipage}
    \begin{minipage}[b]{0.49\linewidth}
        \centering
        \includegraphics[width=1\textwidth]{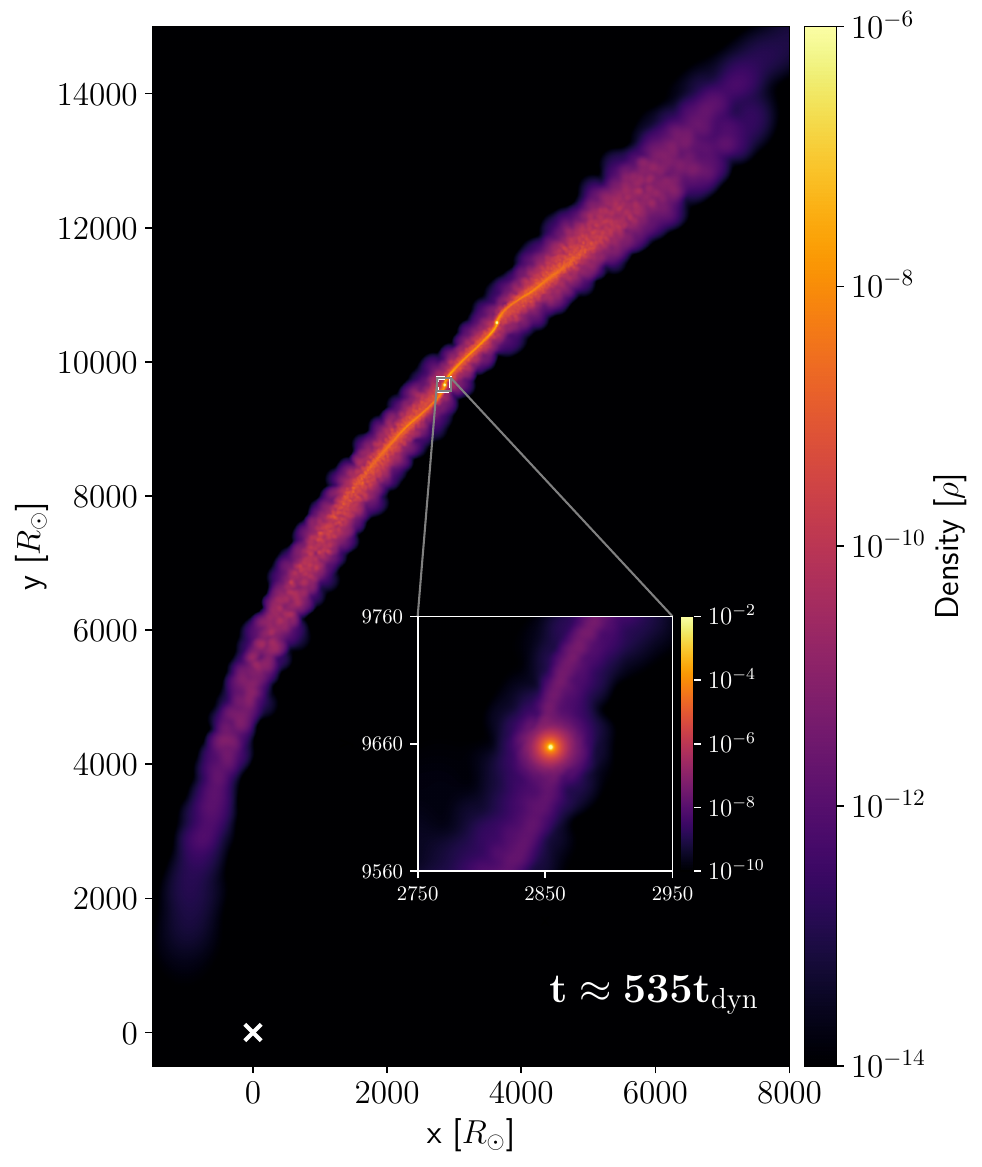}
    \end{minipage}
    \caption{The same as Figure~\ref{fig:deepheadonmovie}, but for Run 2.}
    \label{fig:deepgrazingmovie}
\end{figure*}

\begin{figure*}[htbp]
    \centering
    \begin{minipage}[b]{0.49\linewidth}
        \centering
        \includegraphics[width=1\textwidth]{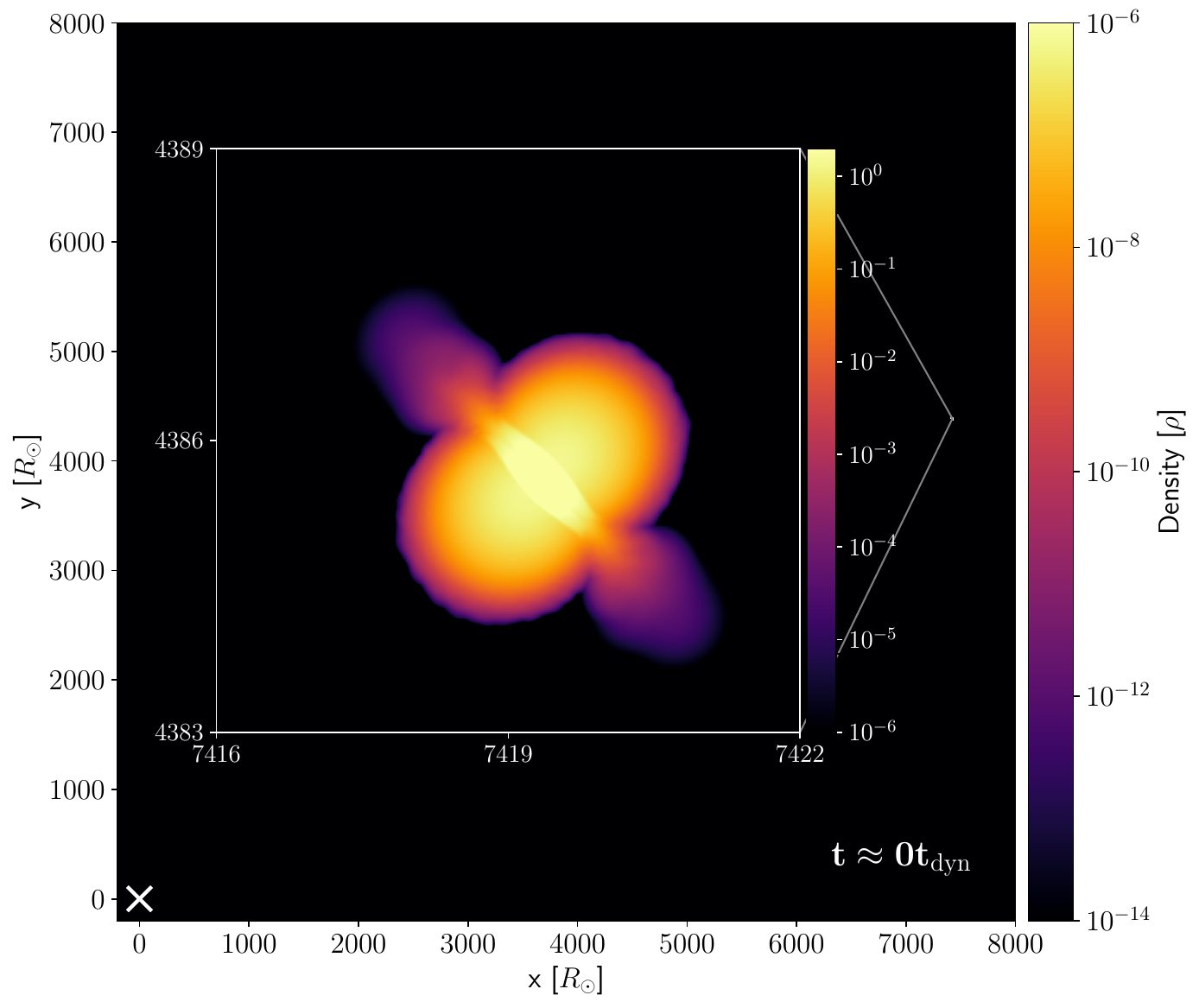}
    \end{minipage}
    \begin{minipage}[b]{0.49\linewidth}
        \centering
        \includegraphics[width=1\textwidth]{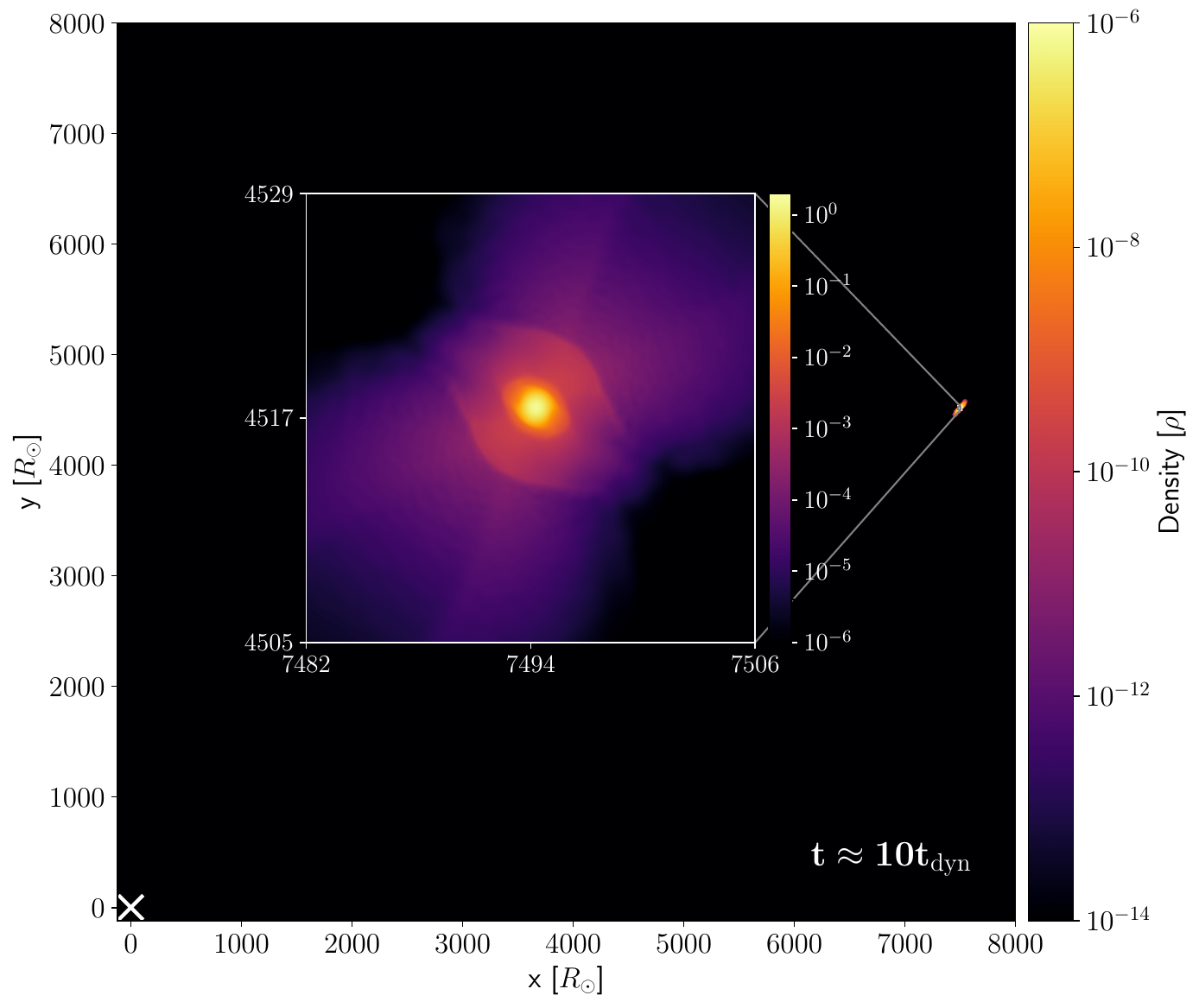}
    \end{minipage}
    \begin{minipage}[b]{0.49\linewidth}
        \centering
        \includegraphics[width=1\textwidth]{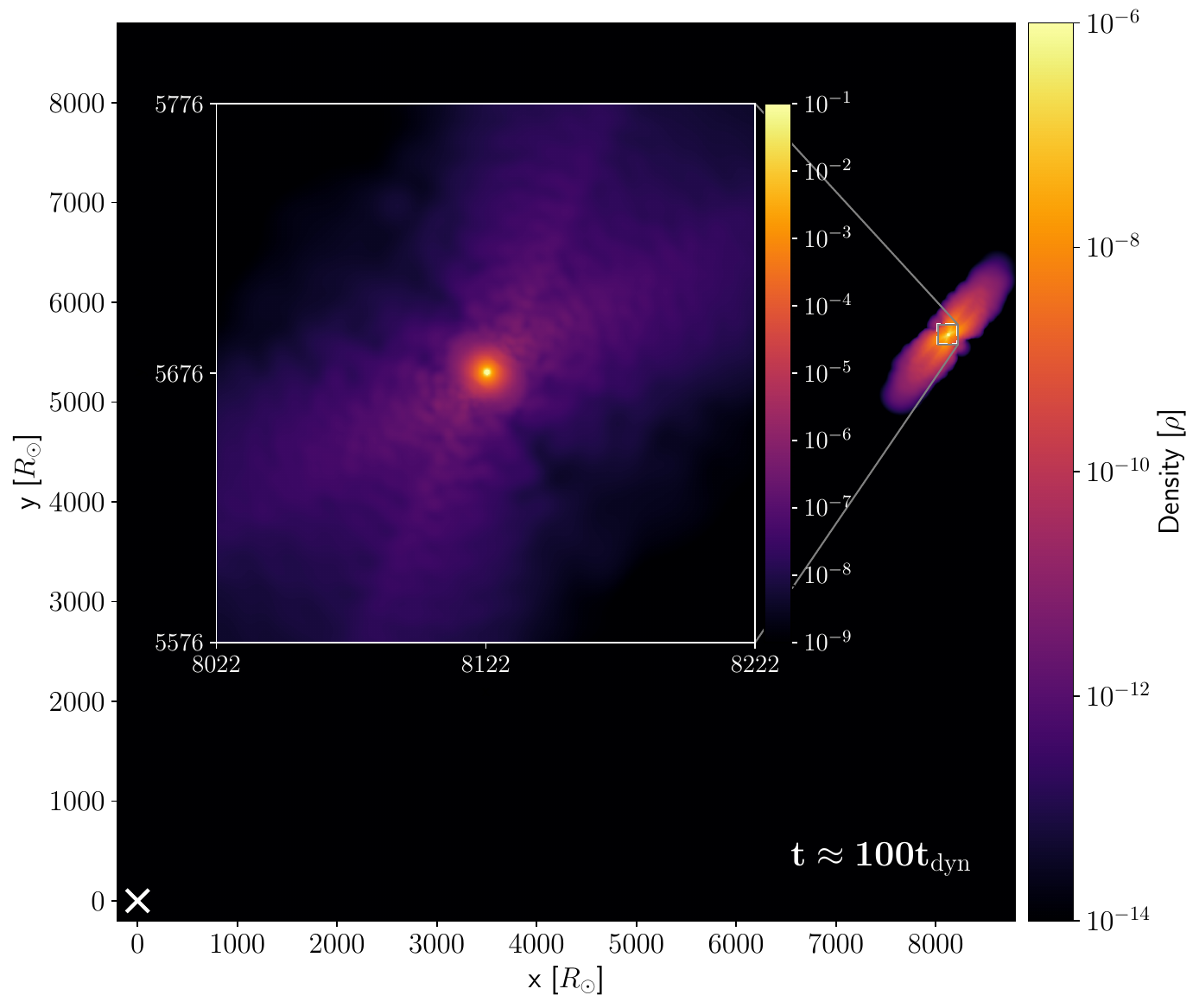}
    \end{minipage}
    \begin{minipage}[b]{0.49\linewidth}
        \centering
        \includegraphics[width=1\textwidth]{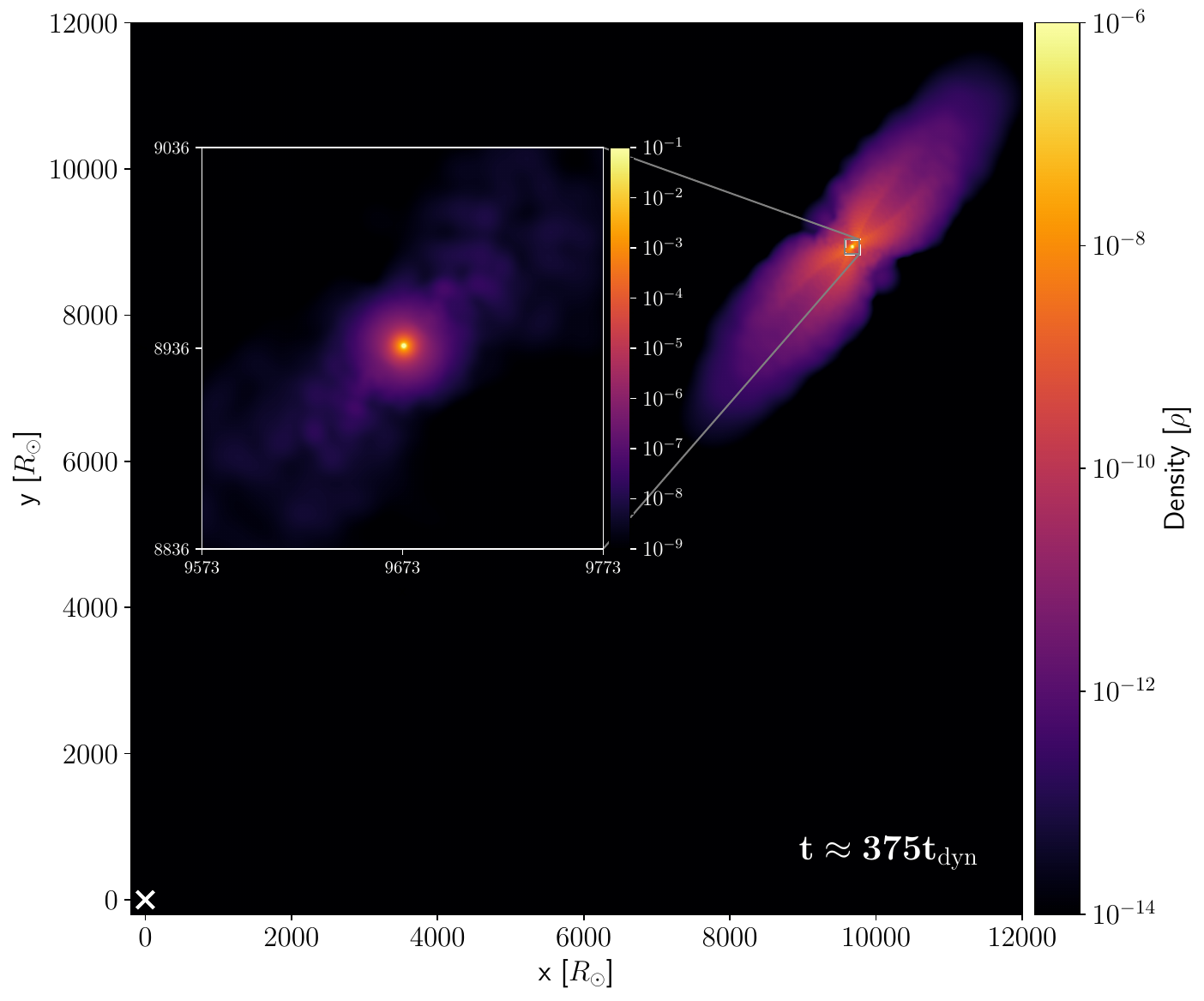}
    \end{minipage}
    \caption{The same as Figure~\ref{fig:deepheadonmovie}, but for Run 3.}
    \label{fig:gentleheadonmovie}
\end{figure*}

\begin{figure*}[htbp]
    \centering
    \begin{minipage}[b]{0.49\linewidth}
        \centering
        \includegraphics[width=1\textwidth]{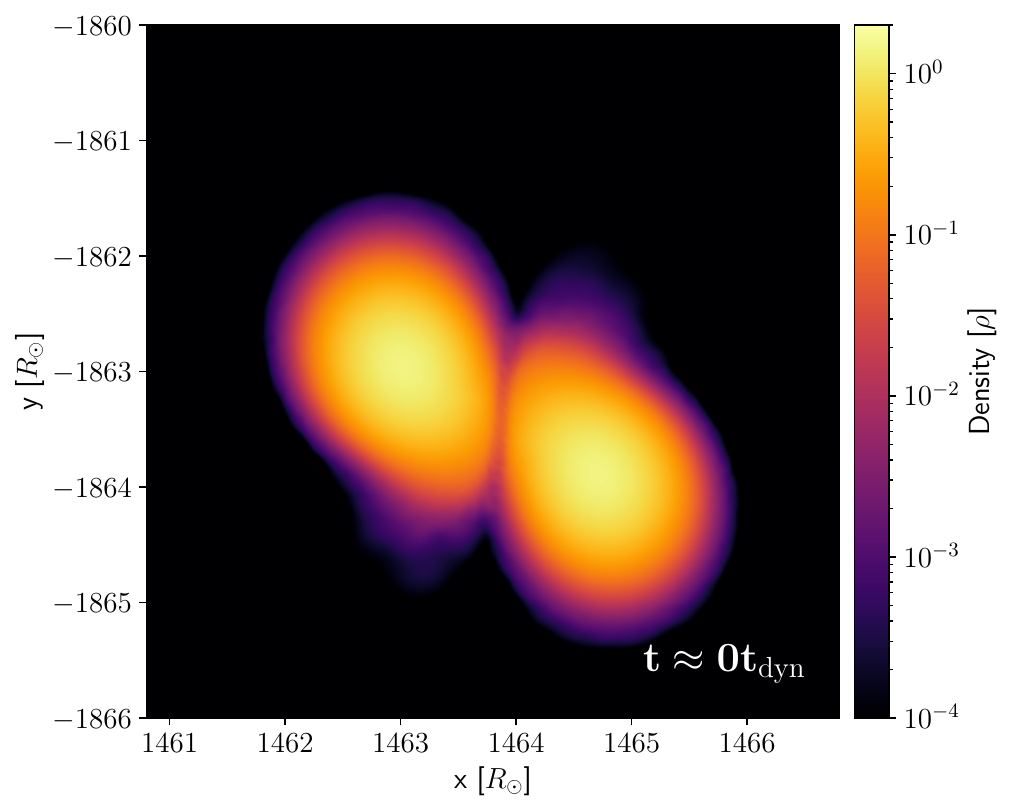}
    \end{minipage}
    \begin{minipage}[b]{0.49\linewidth}
        \centering
        \includegraphics[width=1\textwidth]{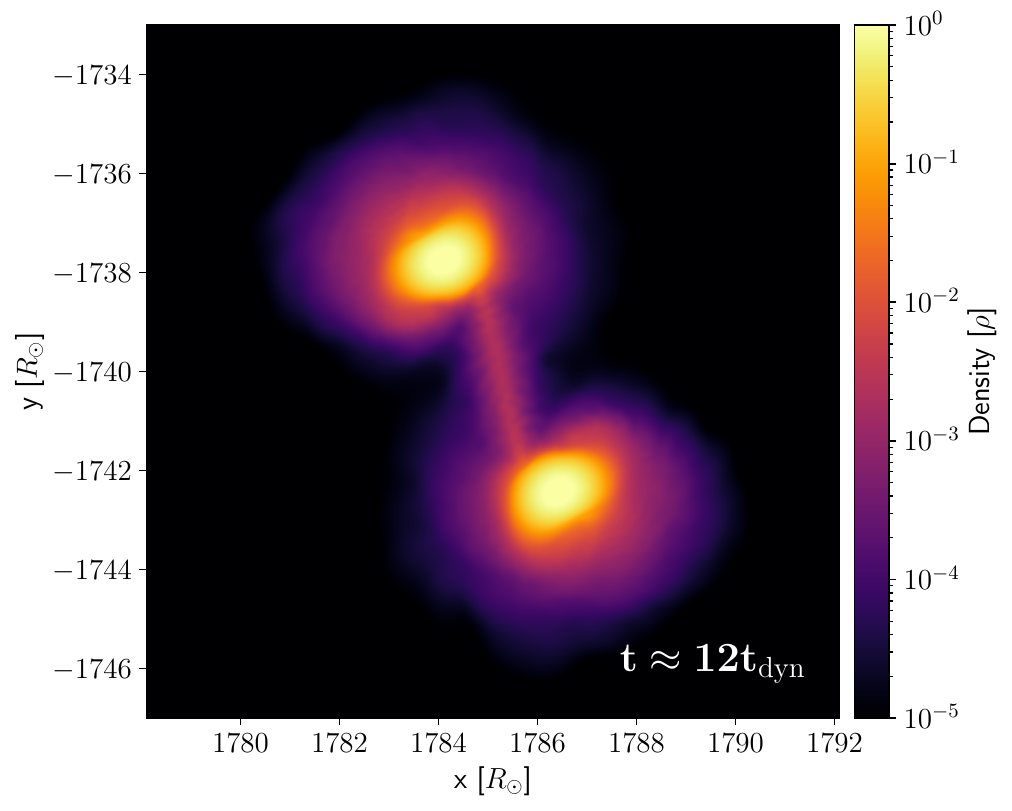}
    \end{minipage}
    \begin{minipage}[b]{0.49\linewidth}
        \centering
        \includegraphics[width=1\textwidth]{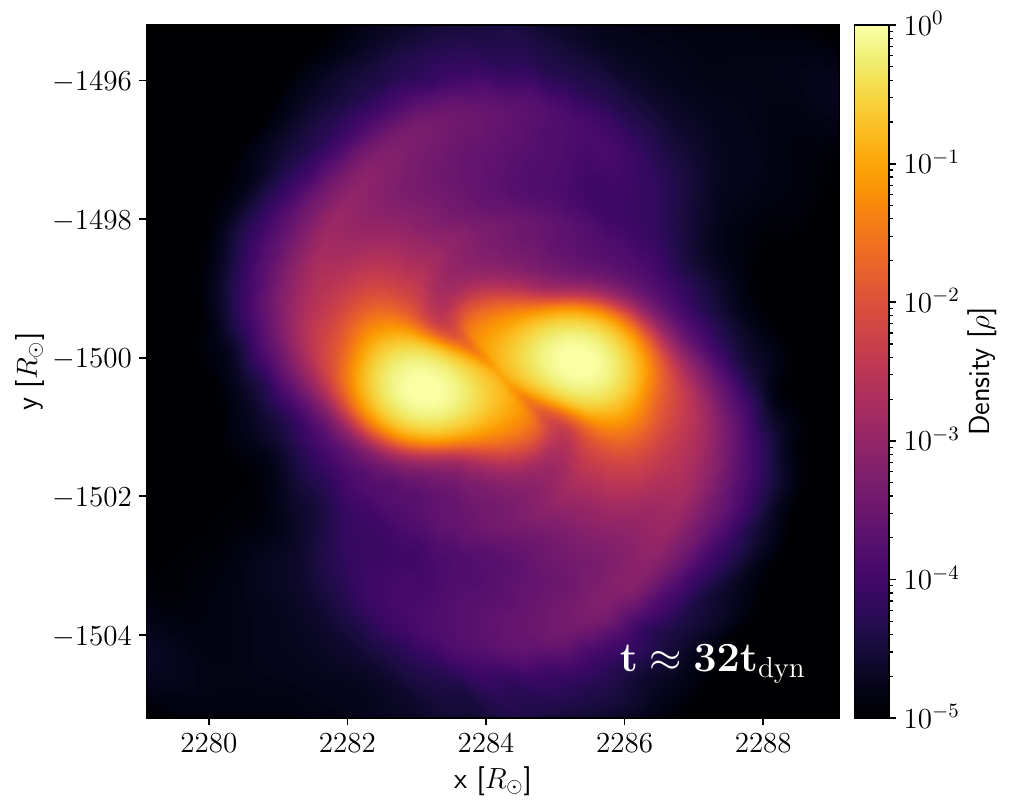}
    \end{minipage}
    \begin{minipage}[b]{0.49\linewidth}
        \centering
        \includegraphics[width=1\textwidth]{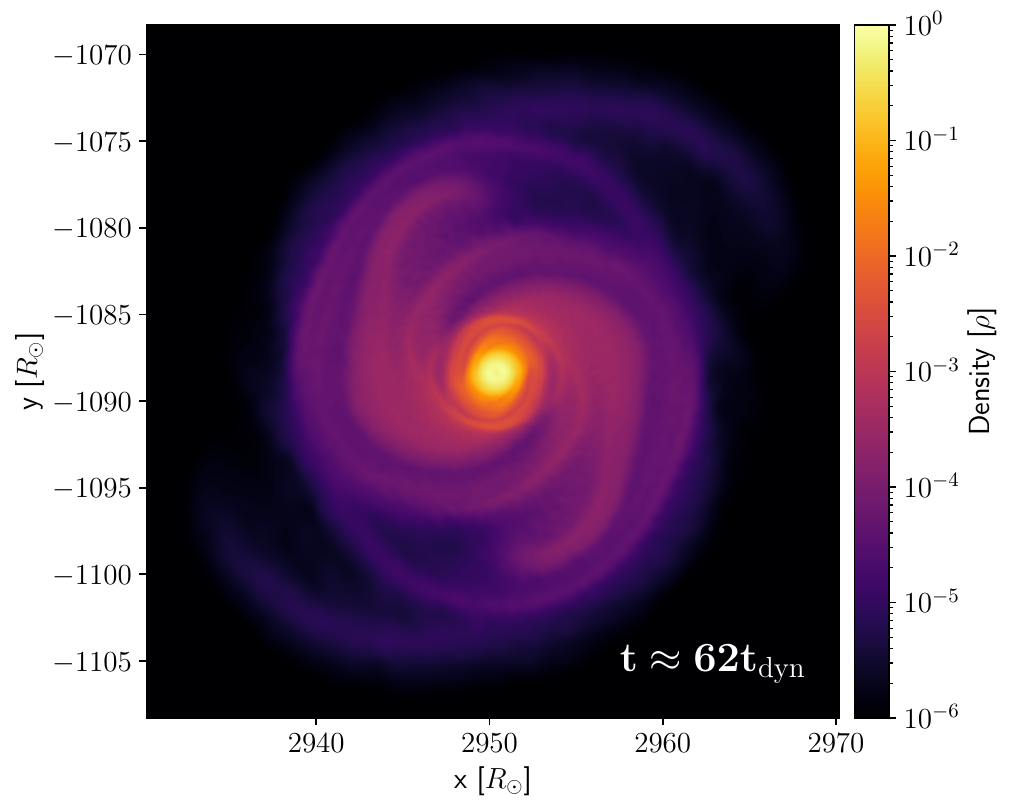}
    \end{minipage}
    \caption{The same as Figure~\ref{fig:gentleheadonmovie}, but for Run 4.}
    \label{fig:gentlegrazingmovie}
\end{figure*}

Figures~\ref{fig:deepheadonmovie}-\ref{fig:gentlegrazingmovie} show the density distribution plot at the $z = 0$ slice (the orbital plane) for Runs 1-4. 
These simulations show that, for both head-on collisions (Run 1 \& 3) and the grazing collision in gentle encounter (Run 4), the stellar collisions produce a single merger remnant. 
In contrast, for the grazing collision in the deep encounter (Run 2), the collision leads to violent perturbations of the two stars but does not result in a merger, leaving behind two remnants instead. 

In addition to the remnants, stellar collisions also generate a significant amount of mass-loss debris, which can fall back onto the SMBH. 
Notably, the resulting ``debris clouds” from these events are distinct from the slender TDE streams typically seen in TDE simulations \citep[e.g.,][]{Evans1989ApJ,Coughlin2016MNRAS,Stone2020SSRv}.

We now examine in detail the properties of the remnants for the four representative simulation runs  for polytropic stars (with $\gamma=5/3$), along with the orbital characteristics of the mass-loss debris and the fallback rate onto the central SMBH.

\subsection{Merger Remnant(s)}
\label{sec:merger remnant}

\begin{figure}[htbp]
    \centering
    \includegraphics[width=\linewidth]{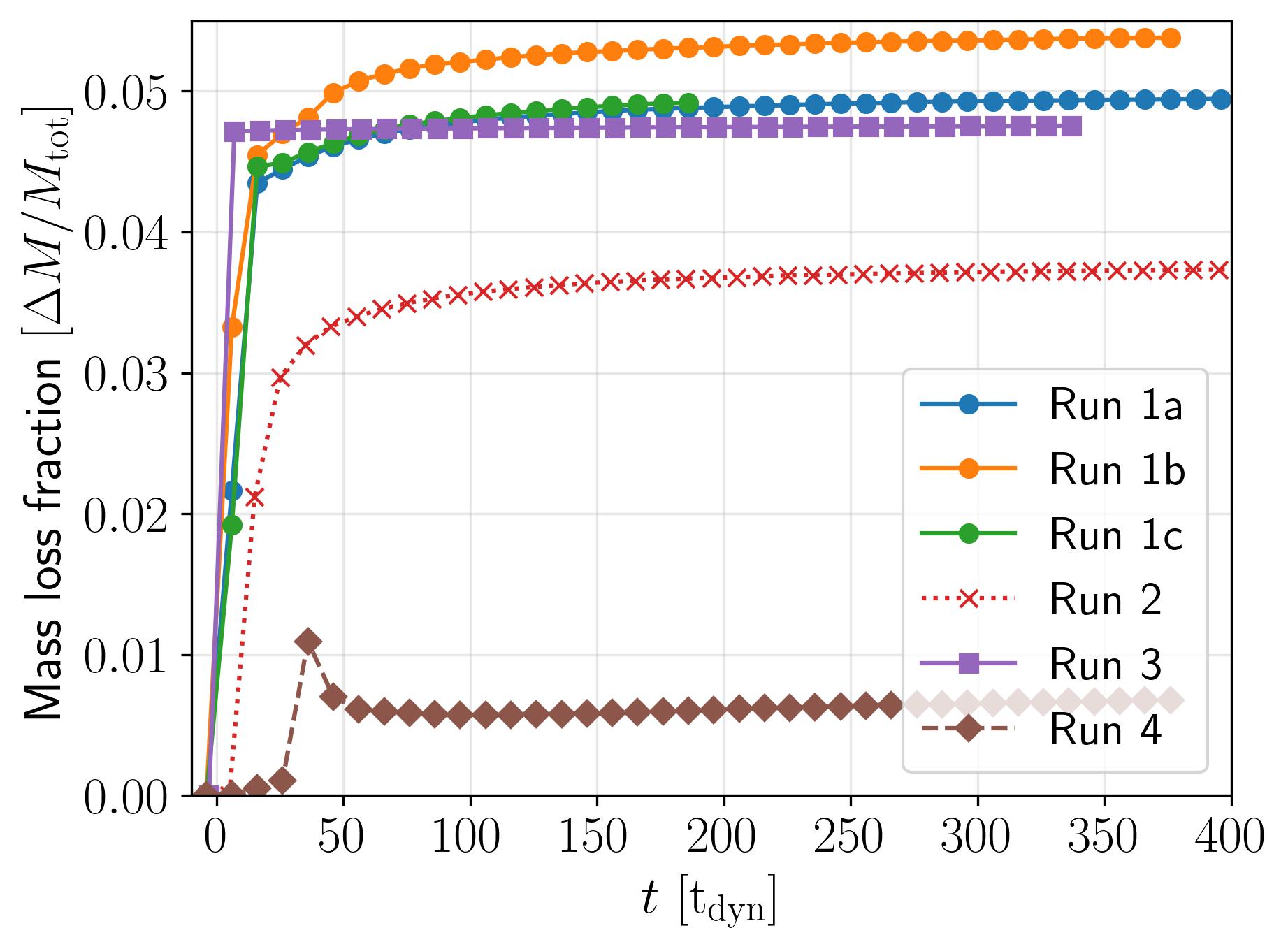}
    \caption{Mass loss fraction as a function of time for Runs 1-4. 
    The lines with the ``$\bullet$'' markers of different colors represent results with different particle numbers ($N=5, 1, 10 \times 10^5$ per star, Runs 1a, 1b, 1c). 
    Note that $t=0$ roughly indicates the moment of  the (first)} collision.
    $M_{\rm tot}$ represents the total mass of the initial binary.
    \label{fig:masslossfrac}
\end{figure}

In Section~\ref{sec:identify}, we describe how we identify the SPH particles that belong to the merger remnant(s).
Figure~\ref{fig:masslossfrac} presents the results of precisely calculating the mass loss fraction using the  iterative Bernoulli method. 
It is evident that the fates of most particles - whether they remain part of the merger remnant or are lost - are determined quickly after the collision. 
This occurs more rapidly in collisions during gentle encounters compared to those in deep encounters  for head-on collisions, as the higher impact velocity in deep encounters results in a more violent interaction among fluid elements, leading to a slower mass loss process.
 For grazing collisions in the gentle encounter (Run 4), the non-monotonic behavior of the mass loss arises from a sequence of successive collisions. 
After the final collision (at $t \gtrsim 40\,t_{\rm dyn}$; see also Figure~\ref{fig:gentle_grazing_sep}), the mass loss fraction gradually settles down to a stable value.

We have conducted a convergence test for head-on collisions in deep encounters using simulations with varying particle numbers (Runs 1a, 1b, 1c). 
The results demonstrate that as the particle number increases, the mass loss fraction quickly converges. 

For head-on collisions in both deep and gentle encounters, the mass loss fraction stabilizes at approximately 5\%.
For grazing collisions, the mass loss fraction is more sensitive to the impact parameter, with less mass loss for larger impact parameters.

\begin{figure*}[htbp]
    \centering
    \begin{minipage}[b]{0.49\linewidth}
        \centering
        \includegraphics[width=\textwidth]{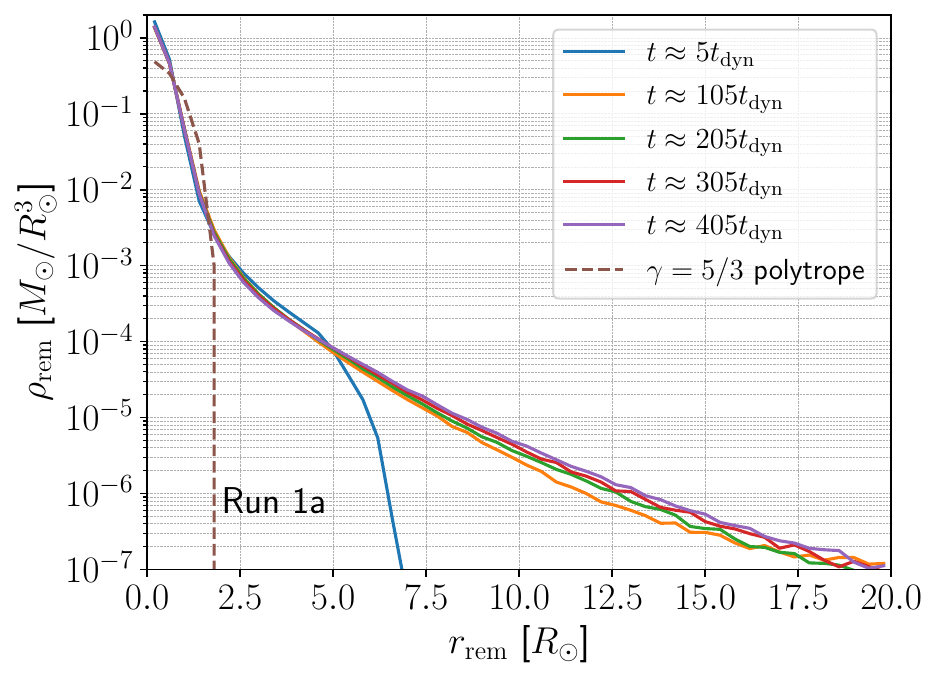}
    \end{minipage}
    \begin{minipage}[b]{0.49\linewidth}
        \centering
        \includegraphics[width=\textwidth]{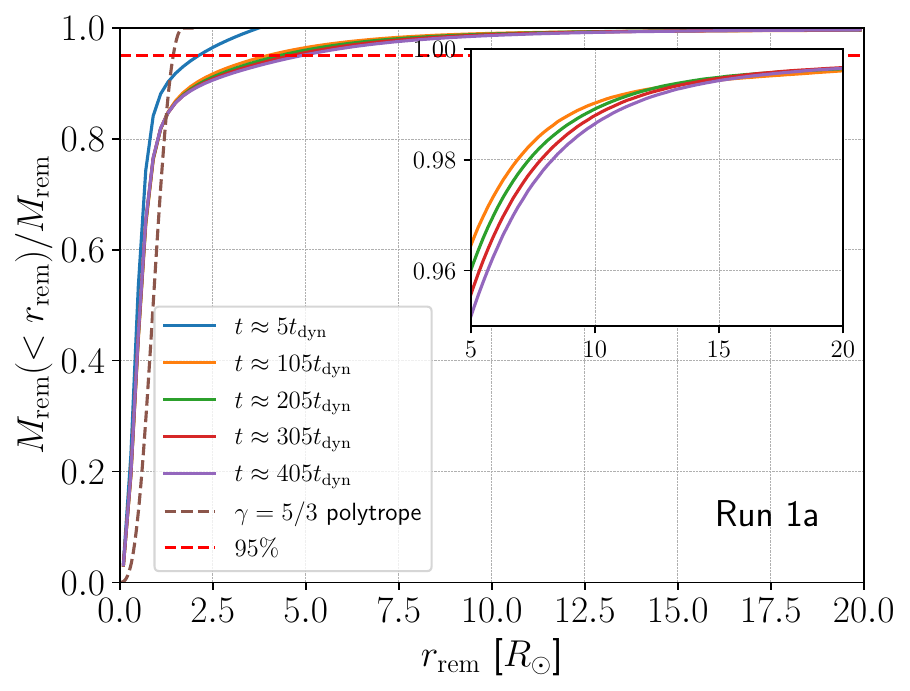}
    \end{minipage}
    \begin{minipage}[b]{0.49\linewidth}
        \centering
        \includegraphics[width=\textwidth]{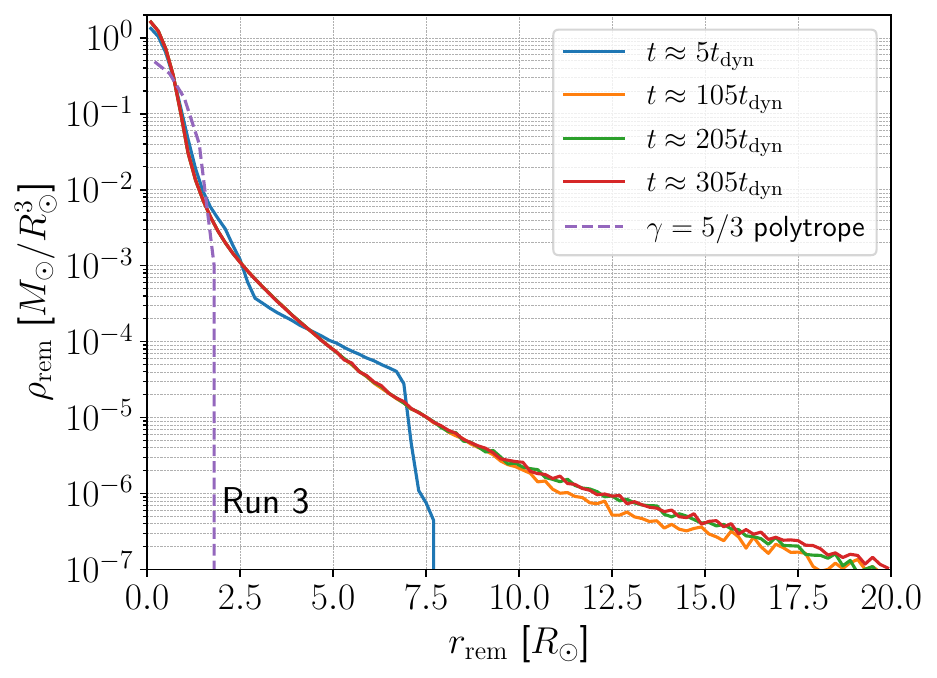}
    \end{minipage}
    \begin{minipage}[b]{0.49\linewidth}
        \centering
        \includegraphics[width=\textwidth]{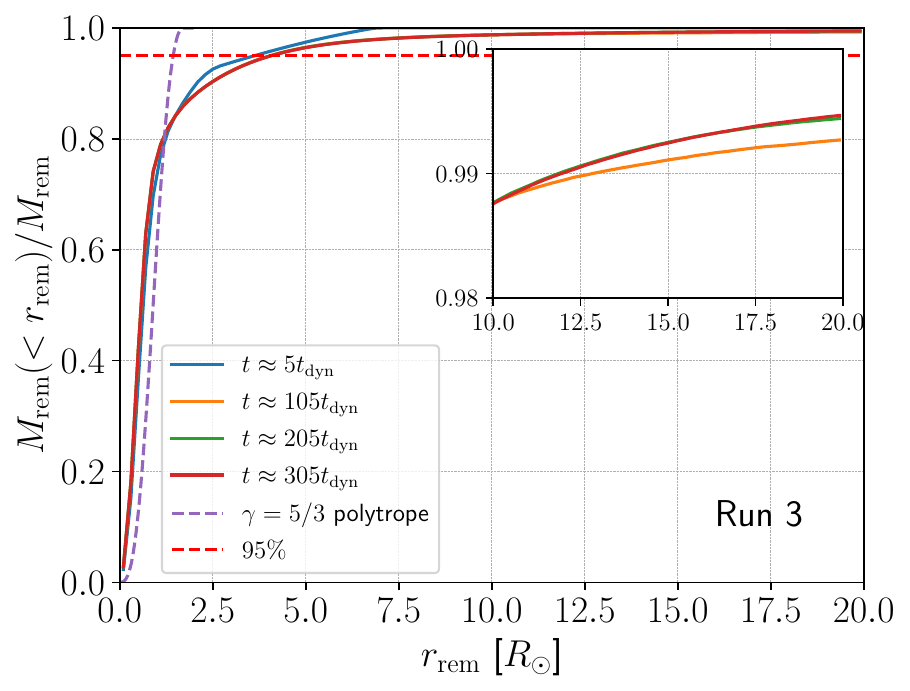}
    \end{minipage}
    \caption{The density profile of the merger remnant (left column) and the fraction of the total mass enclosed within a given radius relative to the total mass of the merger remnant (right column) are shown for Run 1a (upper row) and Run 3 (lower row). 
    The solid lines represent the profiles at different times after the collision, while the dashed line corresponds to a $2 M_\odot$ $\gamma = 5/3$ polytrope with radius determined by the mass-radius relation $R_\star \propto M_\star^{0.8}$, which also approximates the structure of a MS star. 
    The insets in the right panels highlight the mass distribution at the outskirts of the remnant.}
    \label{fig:headon_rho_m}
\end{figure*}

In addition to the mass loss, we are also interested in the structural properties of the merger remnant. 
Clearly, not all particles settle down immediately after the collision, and some asymmetry exists within the merger remnant during the short post-collision phase.
For now, we ignore the effects of this asymmetry. 
Starting from the remnant’s center-of-mass, we calculate the density $\rho_{\rm rem}$ at different radius $r_{\rm rem}$ by dividing the mass and volume of spherical shells. 
We also compute the fraction of the total mass enclosed within a given radius $M_{\rm rem} (<r_{\rm rem})$ relative to the total mass of the merger remnant, as determined using the iterative Bernoulli method. 
Figure~\ref{fig:headon_rho_m} illustrates the structure of the merger remnant following the head-on collision (Runs 1 \& 3).

It can be seen that the main part of the remnant (accounting for 90\% of the total mass) stabilizes within a few $t_{\rm dyn}$, while the remaining 10\% of the mass undergoes small change over hundreds of $t_{\rm dyn}$.
The outermost layers of the envelope continue to expand, and in the case of deep encounter (Run 1), the density in the region $r_{\rm rem} \gtrsim 5 R_\odot$ shows a noticeable increase over time, and 99\% of the total mass is confined within a radius of $15 R_\odot$, which continues to grow slowly. 
In contrast, for the gentle encounter (Run 3), 99\% of the mass is contained within $12 R_\odot$ and remains very stable.

Compared to a typical $2 M_\odot$ MS star (modeled here as a $\gamma = 5/3$ polytrope, with its radius determined by the mass-radius relation $R_\star \propto M_\star^{0.8}$; represented by the dashed line in Figure~\ref{fig:headon_rho_m}), the merger remnant exhibits a more concentrated core and an extremely extended envelope. 
Over 5\% of the total mass is distributed in the envelope at $r_{\rm rem} > 5 R_\odot$.

\begin{figure}[htbp]
    \centering
    \includegraphics[width=\linewidth]{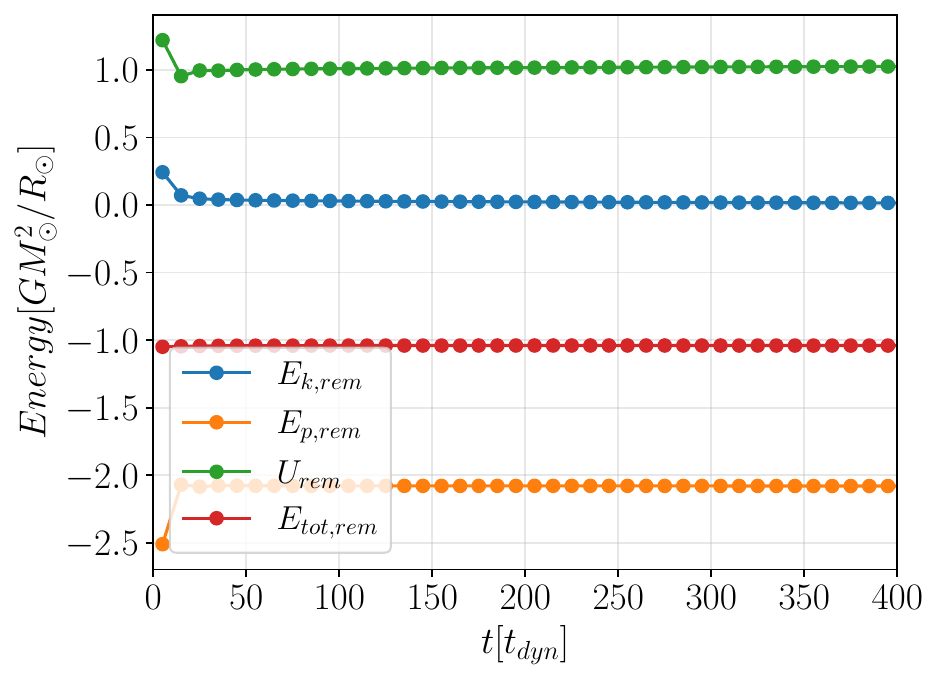}
    \caption{Energy of the merger remnant in Run 1a as a function of time. 
    The different solid lines represent the kinetic energy $E_{k, \rm rem}$ (relative to the remnant center-of-mass), gravitational potential energy $E_{p, \rm rem}$, internal energy $U_{\rm rem}$, and total energy $E_{\rm tot, \rm rem}$.}
    \label{fig:energy}
\end{figure}

\begin{figure}[htbp]
    \centering
    \includegraphics[width=\linewidth]{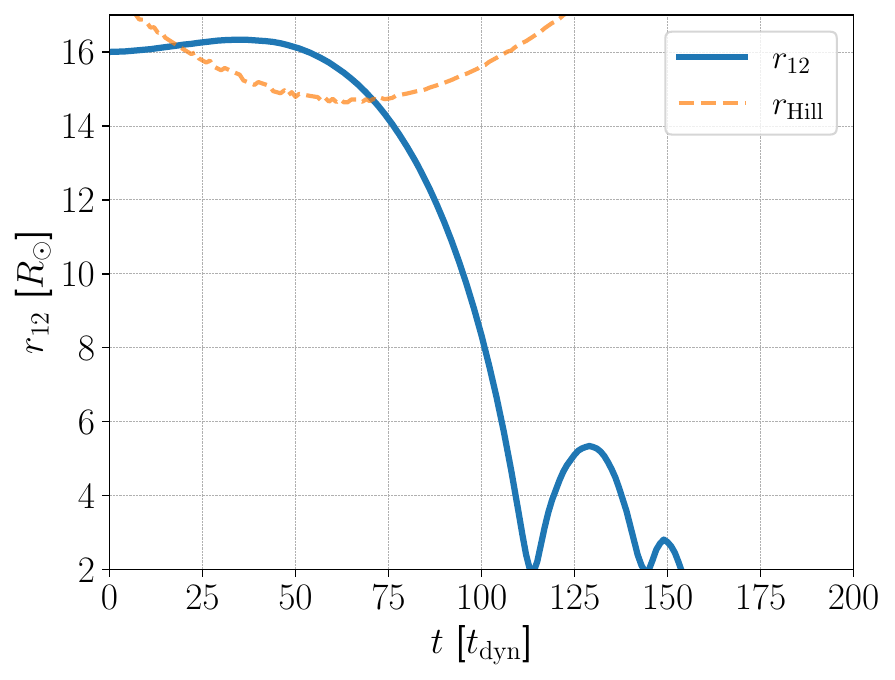}
    \caption{Separation between the two stars in Run 4 as a function of time, where $t=0$ marks the starting point of the simulation. 
    The orange dashed line indicates the Hill radius $r_{\rm Hill}$ of the star.}
    \label{fig:gentle_grazing_sep}
\end{figure}

We can also calculate the total energy of the remnant based on the SPH simulation output. 
Taking the head-on collision in the deep encounter (Run 1a) as an example, we compute the kinetic energy $E_{k, \rm rem}$ relative to the center-of-mass, gravitational potential energy $E_{p, \rm rem}$, and internal energy $U_{\rm rem}$ for all SPH particles belonging to the remnant.
These components were then used to determine the total energy $E_{tot, \rm rem}$ of the remnant.
The results are presented in Figure~\ref{fig:energy}.
Consistent with the mass loss fraction, the various energy components of the star quickly reach a steady state, and the total energy of the remnant remains negative throughout.

For the grazing collisions (Runs 2 \& 4), the larger impact parameters typically prevent the two stars from merging during the first collision. 
In the case of the deep encounter (Run 2), the tidal forces from the SMBH have already disrupted the binary energy-wise. 
After the first collision, the two stars do not collide again but instead undergo a process resembling the standard binary tidal breakup: one star becomes bound to the SMBH’s orbit, while the other is ejected. 
However, both stars have undergone significant perturbations, leaving their structures vastly different from those of typical MS stars.

In contrast, in the gentle encounter (Run 4), Figure~\ref{fig:gentle_grazing_sep} shows that the binary separation remains smaller than the Hill radius $r_{\rm Hill}$ of the star for most of the time, where $r_{\rm Hill}\equiv r_1(M_\star/M_{\rm BH})^{1/3}$ (with $r_1$ the distance of the star to the SMBH). 
The two stars remain gravitationally bound to each other throughout the process. 
Consequently, although the stars do not merge immediately after the first collision, significant dissipation of their relative orbital energy during the collision makes subsequent collisions inevitable. 
These repeated collisions eventually lead to the formation of a single merger remnant. 
In our Run 4, the binary undergoes three collisions before finally merging into a single remnant.

\begin{figure*}[htbp]
    \centering
    \begin{minipage}[b]{0.49\linewidth}
        \centering
        \includegraphics[width=0.95\textwidth]{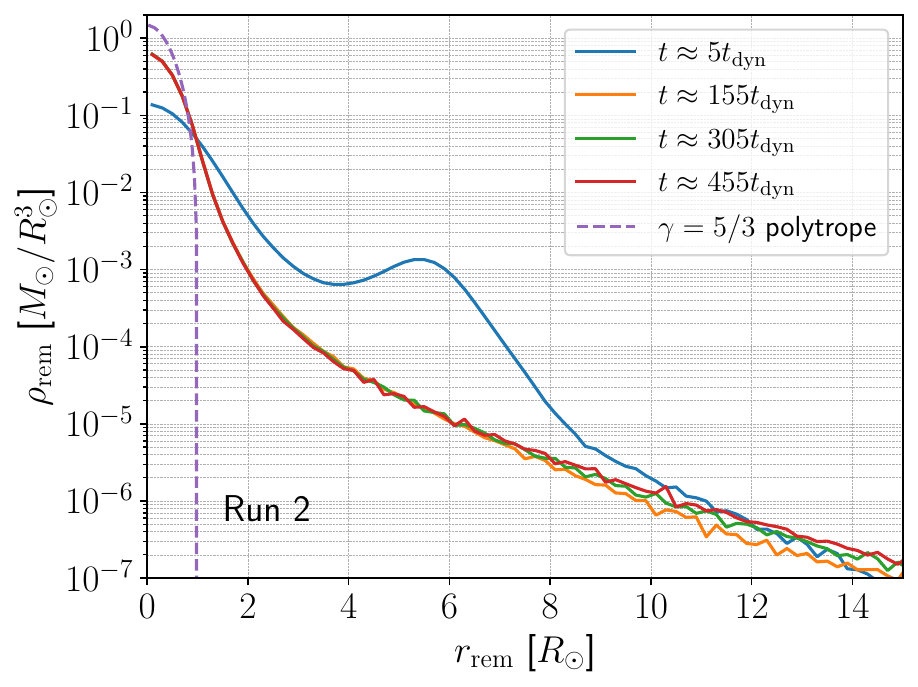}
    \end{minipage}
    \begin{minipage}[b]{0.49\linewidth}
        \centering
        \includegraphics[width=0.95\textwidth]{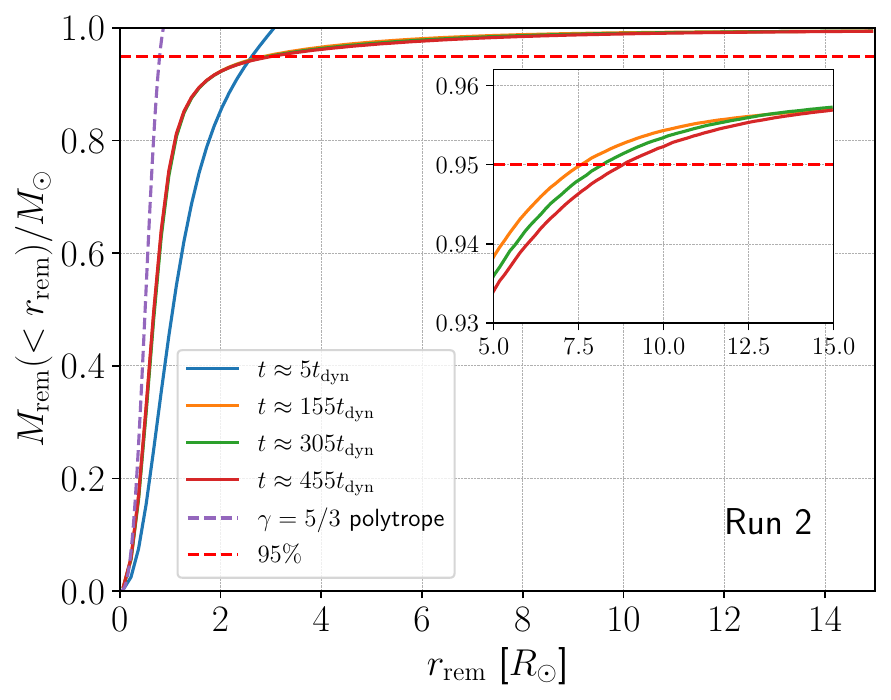}
    \end{minipage}
    \begin{minipage}[b]{0.49\linewidth}
        \centering
        \includegraphics[width=\textwidth]{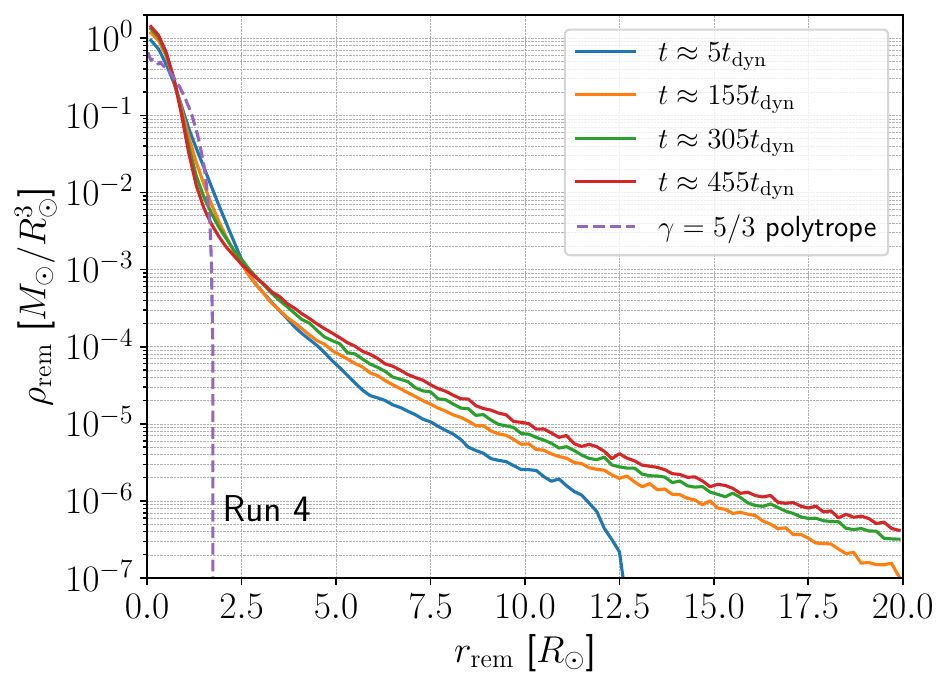}
    \end{minipage}
    \begin{minipage}[b]{0.49\linewidth}
        \centering
        \includegraphics[width=\textwidth]{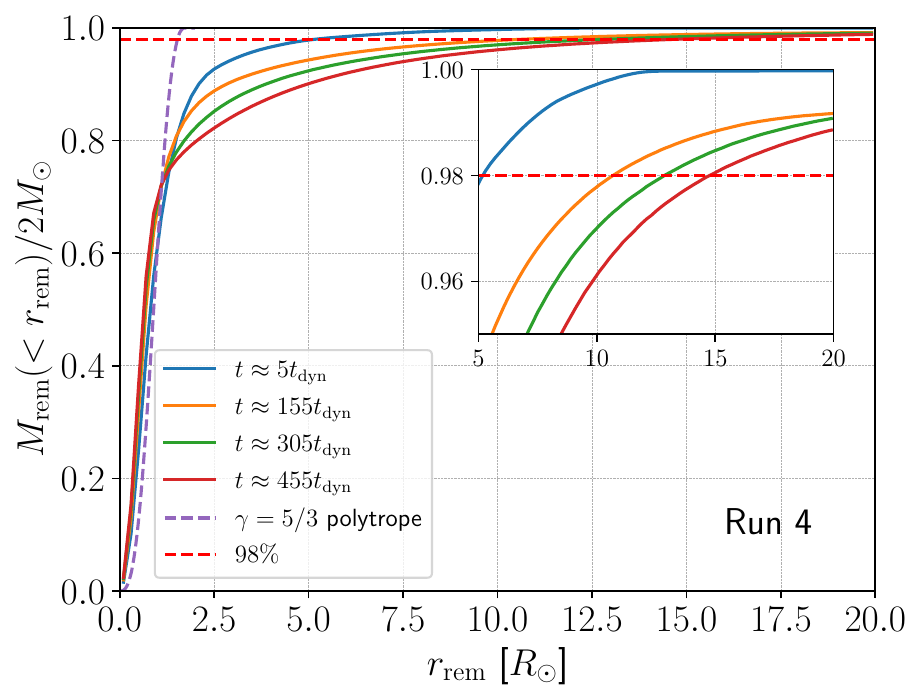}
    \end{minipage}
    \caption{Same as Figure~\ref{fig:headon_rho_m}, but for Runs 2 and 4. 
    Note that the ``bump" in the blue line in the upper left panel represents the other remnant.}
    \label{fig:grazing_rho_m}
\end{figure*}

We also analyze the property of the merger remnant following a grazing collision (Runs 2 \& 4), with the results shown in Figure~\ref{fig:grazing_rho_m}. 
Unlike the cases of head-on collisions, for the deep encounter, we select the remnant that is ultimately bound to the SMBH and compared it to a $1M_\odot$ $\gamma = 5/3$ polytrope. 
 
Similar to the results of head-on collisions (Runs 1 \& 3), the structure of the merger remnant after a grazing collision is also significantly different from that of the original star.
In the deep encounter (Run 2), the remnant resulting from the collision becomes much puffier, with the central density significantly reduced compared to the original star, and a large envelope develops. 
Ignoring the second peak at $5.5 R_\odot$ in the upper left panel of Figure~\ref{fig:grazing_rho_m} (which represents the other remnant), we can observe that more than 90\% of the remnant's mass stabilizes quickly. 
However, as seen in the upper right panel, the remnant continues to expand slowly.

For the merger remnant formed from multiple collisions in the gentle encounter (Run 4), the stellar structure is similar to that of the head-on collision case: a higher central density with a larger, extended envelope. 
However, the stabilization of the merger remnant's structure takes a longer time in this case. 
Even after several hundred $t_{\rm dyn}$, the density distribution still shows changes, and the star's radius continues to increase. 
This is likely due to the higher angular momentum of the merger remnant leading to greater oblateness. 

In Paper I, which assumes perfect inelastic collision when the two stars come into contact, we found that most merger remnants formed in deep encounters, as well as a small fraction of those formed in gentle encounters, are bound to the SMBH if the binary originally follows a parabolic orbit relative to the SMBH. 
We also suggested that one advantage of the merger remnant scenario is that, due to dissipation during the merger process, the remnant stars likely possess very extended envelopes. 
When these remnants approach the SMBH in subsequent orbits, their extended envelopes are highly susceptible to tidal disruption, leading to repeating partial TDEs that can occur multiple times.

Here, we confirm that the merger remnant indeed has a very extended envelope. 
Additionally, our SPH simulations with different impact parameters show that not all collisions result in a single merger remnant (we will discuss how this modifies our results from Paper I in Section~\ref{sec:discussion}). 
Regarding repeating partial TDEs, our simulations indicate that, regardless of whether there is only one merger remnant or two remnants, the collision process makes partial tidal disruption more likely. 
In particular, stars bound to the SMBH after binary breakup will have extended envelopes if they experience grazing collisions, making them more prone to partial disruption.

\subsection{Mass Loss Debris and Fallback}
\label{sec:mass loss}

In Paper I, we suggested the mass loss from stellar collisions induced by binary-SMBH encounters will accrete onto the SMBH, generating accretion luminosity or flares even without a stellar TDE. 
As shown in Figures~\ref{fig:deepheadonmovie}-\ref{fig:masslossfrac}, it is indeed the case that, since the collision velocity is typically comparable to the stellar surface escape velocity (see Figure~9 of Paper I), collisions generally result in a mass loss of approximately 5\%.
This is consistent with the results of stellar collisions in the absence of an SMBH \citep[e.g.,][]{Benz1987ApJ,Lai1993ApJ,Freitag2005MNRAS}.
A key difference between this mass-loss debris and the typical TDE debris is that the former is not produced by tidal forces, and thus does not remain in the same plane or have the very small launch angles characteristic of the TDE debris (which results in an extremely narrow stream).
Instead, the mass-loss debris generated by stellar collisions typically has much larger launch angles, and therefore, in terms of morphology, these debris are more like a ``cloud" than a ``stream".
Similar to TDEs, because our binary is initially on a parabolic orbit relative to the SMBH, only about half of the mass-loss debris will fall back onto the SMBH.

\begin{figure*}[htbp]
    \centering
    \begin{minipage}[b]{0.49\linewidth}
        \centering
        \includegraphics[width=\textwidth]{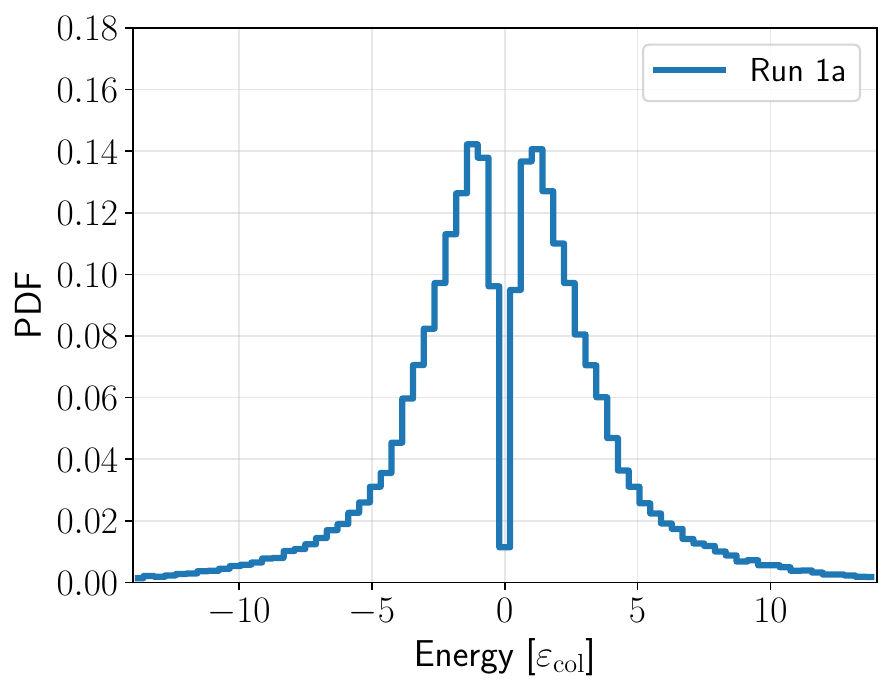}
    \end{minipage}
    \begin{minipage}[b]{0.49\linewidth}
        \centering
        \includegraphics[width=\textwidth]{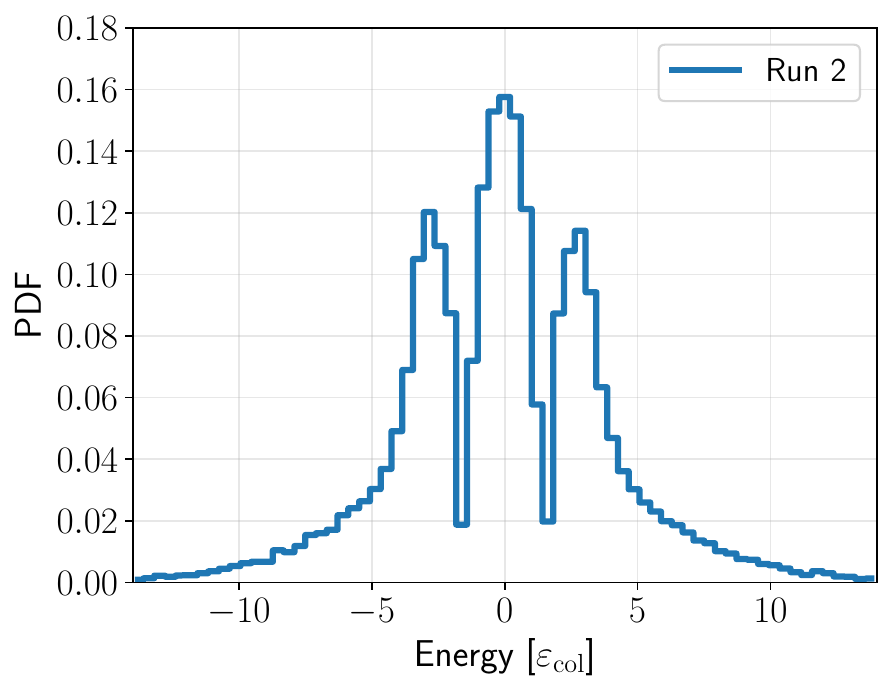}
    \end{minipage}
        \begin{minipage}[b]{0.49\linewidth}
        \centering
        \includegraphics[width=\textwidth]{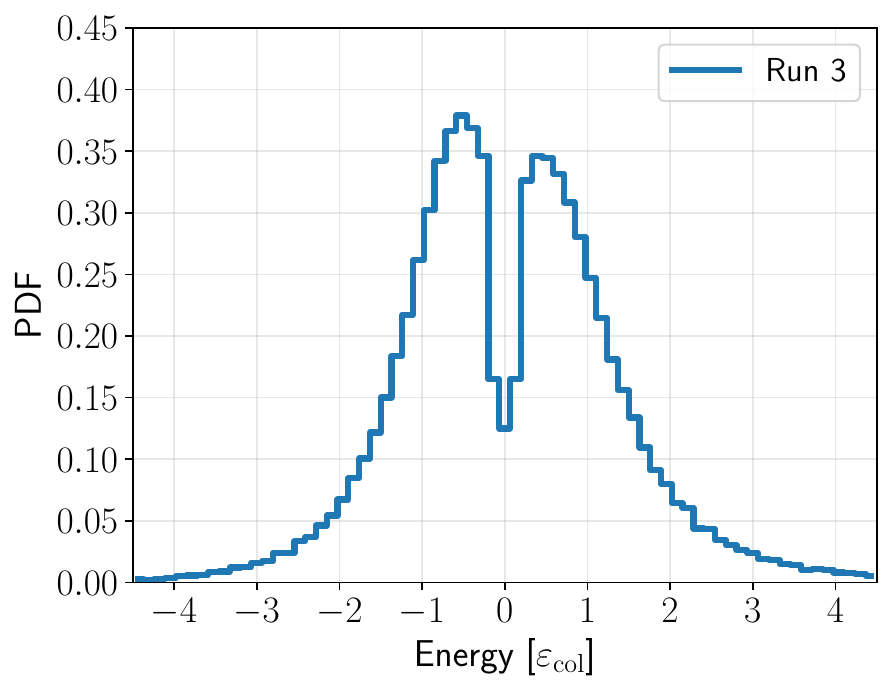}
    \end{minipage}
    \begin{minipage}[b]{0.49\linewidth}
        \centering
        \includegraphics[width=\textwidth]{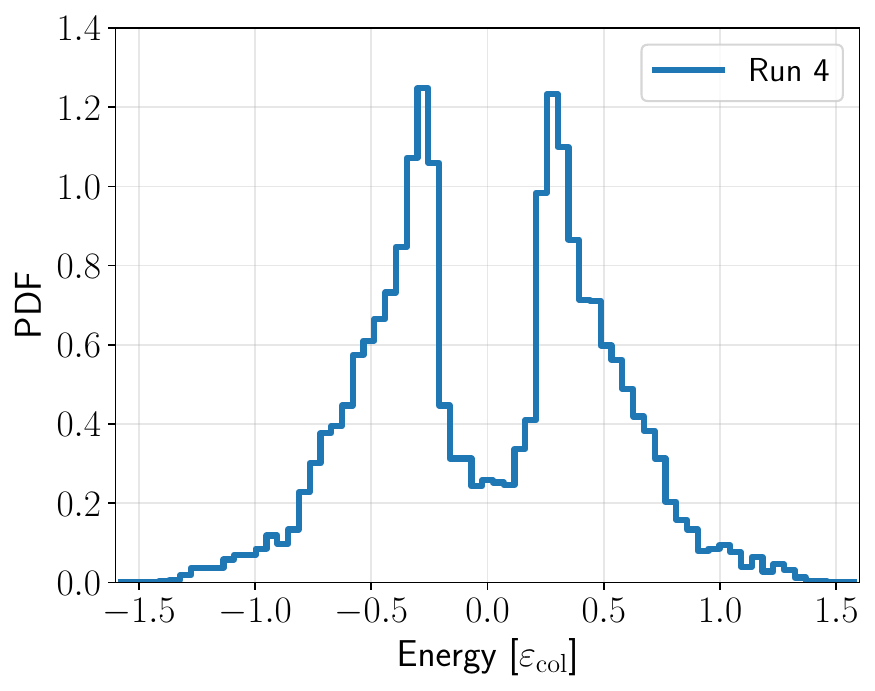}
    \end{minipage}
    \caption{Probability density function (PDF) of the orbital energy of the mass-loss debris relative to the SMBH after collisions. 
    The results are scaled by $\varepsilon_{\rm col}$, as defined in Eq.~(\ref{eq:ecol}).}
    \label{fig:massloss_energy}
\end{figure*}

We begin by analyzing the orbital energy of the mass-loss debris relative to the SMBH, with the results shown in Figure~\ref{fig:massloss_energy}.
It is clear that the energy distribution of the debris produced by stellar collisions is quite different from that of the TDE debris. 
Here, we scale the orbital energy of the debris relative to the SMBH by the tidal energy of the SMBH at pericenter during a deep encounter:
\begin{equation}
    \varepsilon_{\rm col} \equiv \frac{G M_{\rm BH} R_\star}{r_p^2}=\frac{GM_{\rm BH} R_\star}{{r_{\rm tide}^\star} ^2} \beta_\star^2.
\label{eq:ecol}
\end{equation}
Although the collision in a deep encounter does not occur exactly at the pericenter, and this energy scale is less justified in the case of a gentle encounter, Figure~\ref{fig:massloss_energy} shows that Eq.~(\ref{eq:ecol}) works well when describing the energy distribution, especially in terms of the peak value.

From Figure~\ref{fig:massloss_energy}, we see that for collisions that result in a single remnant (Runs 1a, 3, and 4), the debris energy distribution exhibits a bimodal structure with a relatively long tail.
The bimodal feature arises because the orbital energy of the merger remnant relative to the SMBH is nearly zero, so most particles with nearly zero orbital energy remain bound to the merger remnant.
The long tail indicates that many debris particles have relatively high launch velocities relative to the merger remnant.
This is due to the shock waves and fluid forces present during the collision, allowing some debris to achieve higher energies relative to the merger remnant.
 For a grazing collision in a deep encounter that results in two perturbed stars (Run 2; upper-right panel in Figure~\ref{fig:massloss_energy}), the energy distribution exhibits a triple-peaked structure in addition to the long tail. 
This can be interpreted in a similar manner as the bimodal cases, where debris particles around the dips are bound to the corresponding stellar remnants. 
In this case, however, the remnants themselves have non-zero orbital energies—approximately symmetric and opposite in sign—corresponding to the two dips near $\varepsilon_{\rm col} \sim \pm 2$.
We also see that in our representative examples, the collisions in deep encounter (Run 1a , 2) results in a wider energy distribution compared to the collision in the gentle encounter (Run 3 , 4).

\begin{figure*}[htbp]
    \centering
    \begin{minipage}[b]{0.49\linewidth}
        \centering
        \includegraphics[width=\textwidth]{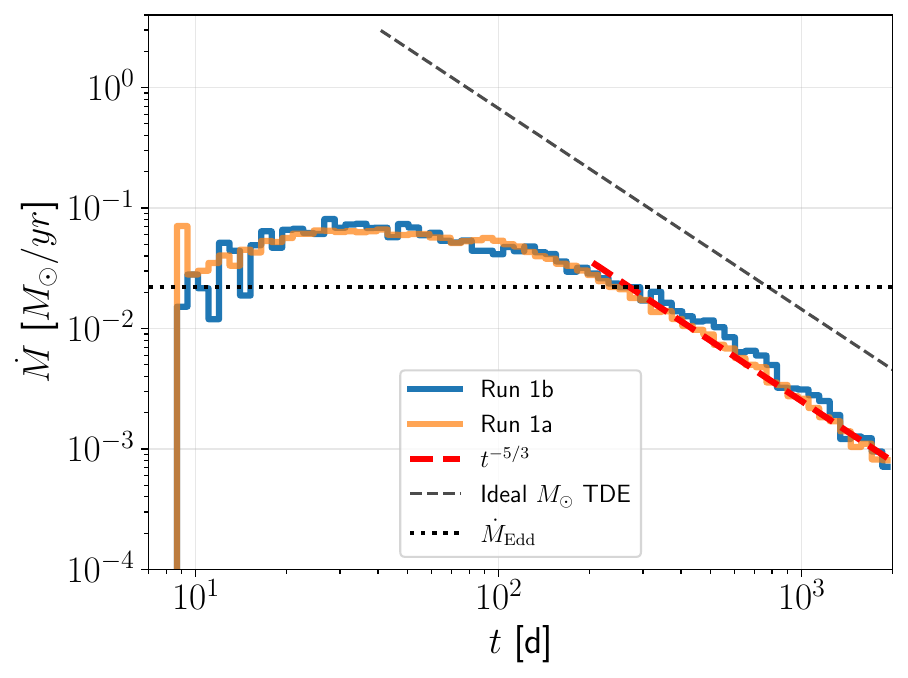}
    \end{minipage}
    \begin{minipage}[b]{0.49\linewidth}
        \centering
        \includegraphics[width=\textwidth]{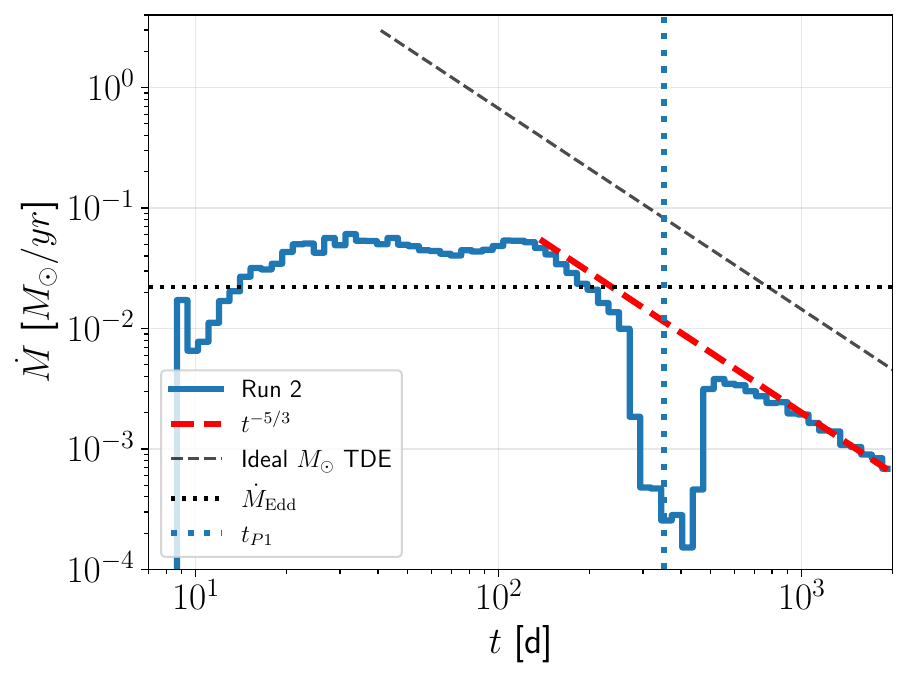}
    \end{minipage}
    \begin{minipage}[b]{0.49\linewidth}
        \centering
        \includegraphics[width=\textwidth]{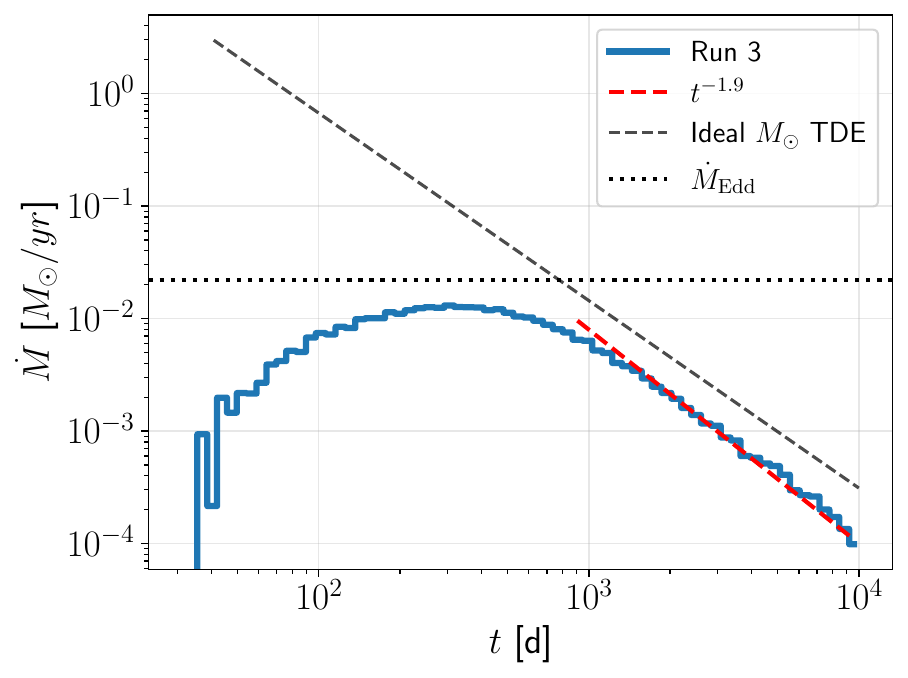}
    \end{minipage}
    \begin{minipage}[b]{0.49\linewidth}
        \centering
        \includegraphics[width=\textwidth]{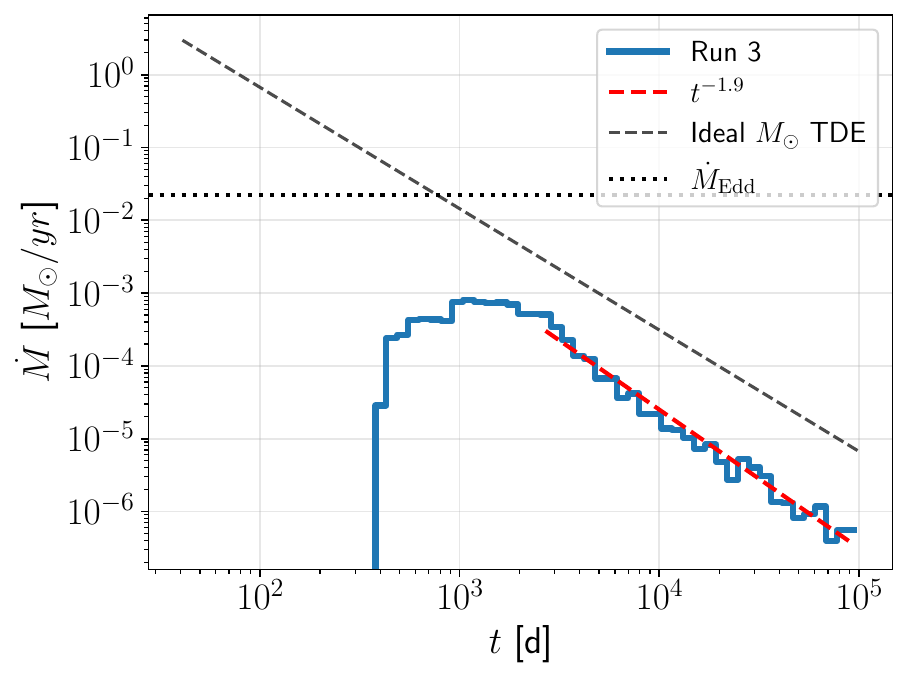}
    \end{minipage}
    \caption{Fallback accretion rate, $\dot{M}(t)$, onto the SMBH. 
    The different colored lines  in the upper left panel represent the results for different particle numbers ($N=5\times10^5$ and $10^5$ per star for Run 1a and 1b).
    To convert to physical units, we  have set $M_{\rm BH}=10^6M_\odot$, $M_\star=1M_\odot$ and $R_\star=1R_\odot$.
    The black dashed line shows the standard mass fallback rate for a $1M_\odot$ star undergoing a TDE, while the red dashed line represents a power-law curve that matches the late-time fallback rate. 
    The dotted  horizontal line indicates the Eddington limit for a $10^6 M_\odot$ black hole, assuming an accretion efficiency of $\epsilon = 0.1$. 
    The blue dotted vertical line in the upper right panel marks the moment when the merger remnant returns to the pericenter of its orbit around the SMBH.
    Here, $t=0$ corresponds to the moment of stellar collision.}
    \label{fig:fallbackrate}
\end{figure*}

For the mass-loss debris that falls back onto the SMBH, we are interested in their fallback rate, $\dot{M}(t)$, as it is strongly correlated with the observable light curve. 
Using the method for calculating $\dot{M}(t)$ described in Section~\ref{sec:CalculationMdot}, we compute the fallback time for each SPH particle identified as mass loss. 
The results are shown in Figure~\ref{fig:fallbackrate}.

In this figure, we also compare the fallback rates for simulations with different particle numbers ($N=5, 1 \times 10^5$ per star) and contrast them with the standard mass fallback rate from the TDE of a $1M_\odot$ star, as well as the Eddington limit for a $10^6 M_\odot$ SMBH, assuming an accretion efficiency of $\epsilon = 0.1$.

The fallback rate curves of the mass-loss debris differ significantly from the standard TDE curve. 
These curves exhibit a slow rise over an extended period, remain relatively steady, and then follow a power-law decline similar to that of a TDE. 
 For the grazing collision in a deep encounter, as previously discussed, debris with specific orbital binding energy similar to that of the bound remnant tends to remain gravitationally bound and effectively becomes part of the remnant. 
This is reflected in the dip appearing around $\sim 250$ days in the curve (Figure~\ref{fig:fallbackrate}, the upper right panel), which coincides with the time when the remnant returns to the pericenter of its orbit around the SMBH.
The relationship between the curve shape and $\beta_b$ is not very clear.
The peak fallback rate is much lower than that of a TDE, as the total fallback mass is only a few percent of the fallback mass in a TDE for a star of the same mass. 
However, the minimum fallback time for the debris is significantly shorter compared to the TDE case, particularly for deep encounters (the upper row). 
In gentle encounters, this effect is less pronounced, as the collision occurs long after the pericenter passage.
Despite comparable total mass loss, there is a significant difference in the magnitude of the peak fallback rates. 
In the deep encounter (Run 1 , 2), the collision occurs very close to the pericenter, causing the debris to fall back onto the SMBH more quickly. 
This results in a substantially higher peak rate, which remains above the Eddington limit for several hundred days. 
In contrast, for the gentle encounter (Run 3 , 4), the fallback occurs much more slowly, and the curve consistently stays below the Eddington limit.

After the mass-loss debris falls back to the pericenter around the SMBH,
we expect it to undergo orbital circularization and disk formation, and eventually accrete onto the SMBH, analogous to that of TDE debris \citep{Bonnerot2021SSRv}.

However, since the ``cloud" of debris produced by stellar collisions differs significantly from the TDE stream, the energy dissipation process may also differ substantially.
A quantitative calculation of the light curve of BH accretion caused by stellar collisions is beyond the scope of this paper. 
Nevertheless, we can provide some qualitative insights. 
Due to the large width of the debris ``cloud", the compression effects during fallback onto the SMBH are expected to be more pronounced compared to the TDE streams, leading to greater energy dissipation from nozzle shocks. 
The inclination range of the debris produced by collisions is also broader than in TDEs. 
As a result, the density at pericenter may be lower than in the case of coplanar debris, but the intersection of orbits with different inclinations near pericenter could result in additional energy dissipation.

\begin{figure*}[htbp]
    \centering
    \begin{minipage}[b]{0.49\linewidth}
        \centering
        \includegraphics[width=\textwidth]{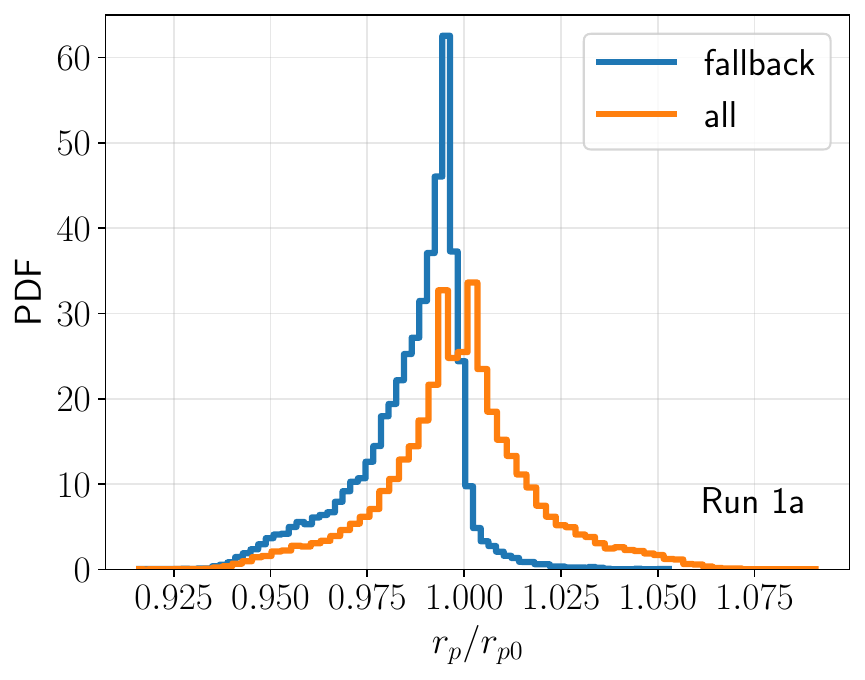}
    \end{minipage}
    \begin{minipage}[b]{0.49\linewidth}
        \centering
        \includegraphics[width=\textwidth]{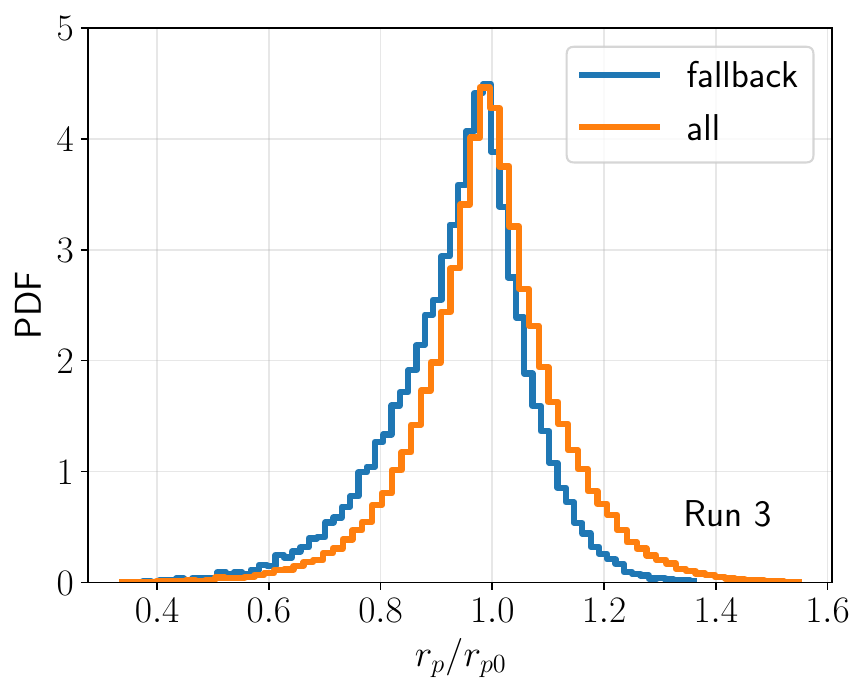}
    \end{minipage}
    \caption{Probability density function (PDF) of the pericenter distance ($r_p$) of the mass-loss debris relative to the SMBH in Runs 1a and 3. 
    The results are scaled by the initial pericenter distance of the binary orbit relative to the SMBH $(r_{p0})$. 
    The orange line represents all mass-loss debris, while the blue line represents the subset of debris that will fall back onto the SMBH.}
    \label{fig:rp}
\end{figure*}

We can estimate the approximate size of the final disk by calculating the pericenter distances of the mass-loss SPH particles, based on their angular momenta (relative to the SMBH) after the  head-on collision s (Runs 1a, 3). 
The results are shown in Figure~\ref{fig:rp}.

We see that after the deep encounter head-on collision (Run 1a), the angular momentum distribution of the debris exhibits only about $5\%$ variation compared to the initial angular momentum, with the fallback debris primarily concentrated in the region where just inside $r_{p0}$ (the initial pericenter distance of the binary). 
In contrast, during the gentle encounter head-on collision, the angular momentum distribution of the debris varies by about $40\%$ compared to the initial value.
The difference in these distributions arises because deep encounter collisions occur very close to the SMBH, where the orbital velocity is much greater than the velocity of the debris relative to the remnant. 
As a result, the debris angular momentum remains largely unaffected. 
On the other hand, gentle encounter collisions occur far from $r_{p0}$, at the location where the orbital velocity is already relatively small, making the angular momentum more susceptible to change.

\section{Results: \texttt{MESA} stars}
\label{sec:results_MESA}
The $\gamma=5/3$ polytrope is considered in section~\ref{sec:results_poly} underestimates the central density concentration of solar-type MS stars. 
As the hydrodynamics of stellar collisions is highly sensitive to the initial stellar mass distribution \citep[e.g.,][]{Lai1993ApJ}, we perform a comparative set of simulations using a more realistic $M=M_\odot$ MS star model generated by the stellar evolution code \texttt{MESA}. 
For simplicity and to maintain a clear comparison with the polytropic case presented in section~\ref{sec:results_poly}, we focus on head-on and grazing collisions in the deep encounter regime.

The overall collisional dynamics share many qualitative similarities with the polytropic results. 
The most striking difference, however, lies in the outcome of the grazing encounter. 
In the polytropic case, the two stars experience a single pericenter passage and continue on unbound hyperbolic trajectories as two separate, highly perturbed stars. 
For the \texttt{MESA} stars, we find the different outcome is driven by the initial collision parameters. A comparison of Table~\ref{tab:simulations} shows that the relative velocity for the \texttt{MESA} star encounter (Run 6) is lower than for the polytropic case (Run 2). This lower kinetic energy allows the stars to become gravitationally bound.
Consequently, they are unable to escape each other and are quickly drawn back for a second, decisive encounter that leads to a full merger into a single remnant. 
This process of tidal capture followed by a merger is analogous to the behavior seen in the gentle grazing encounter of polytropic stars depicted in Figure~\ref{fig:gentle_grazing_sep}.

We now analyze three key diagnostics as in Section~\ref{sec:results_poly}: the mass loss fraction, the post-merger remnant structure, and the fallback accretion rate of mass-loss debris.

\begin{figure}[htbp]
    \centering
    \includegraphics[width=\linewidth]{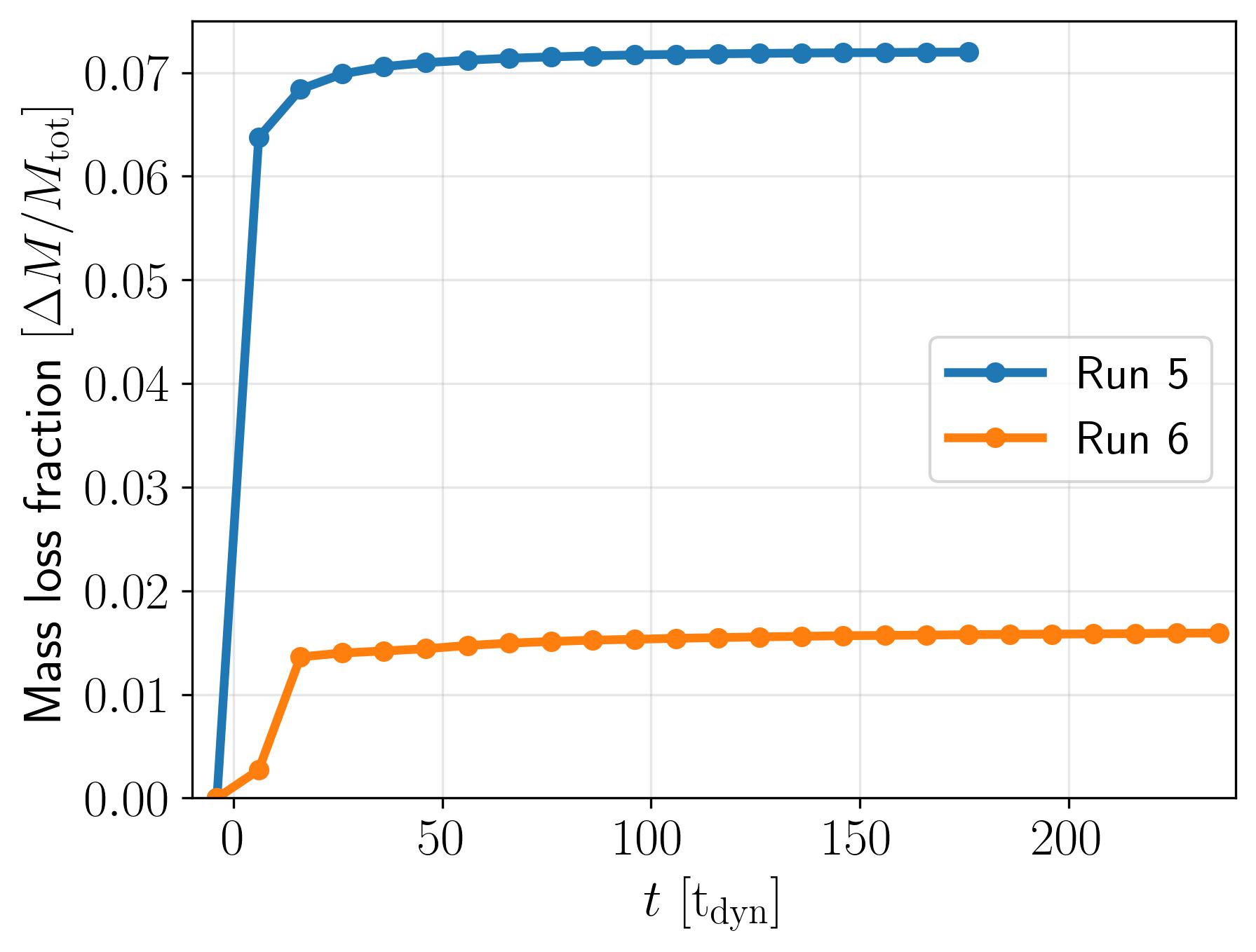}
    \caption{Same as Figure~\ref{fig:masslossfrac}, but for Runs 5 and 6.}
    \label{fig:MESAmasslossfrac}
\end{figure}

Figure~\ref{fig:MESAmasslossfrac} presents the mass loss fractions calculated using the iterative Bernoulli method for Runs 5 and 6.  
Similar to the polytropic cases, the mass loss fractions for \texttt{MESA} stars also stabilize shortly after the collision.  
The stellar structures are different for the two collision types. 
The head-on collision (Run 5) ejects more mass than its polytropic equivalent. 
This is because the highly concentrated cores of the \texttt{MESA} stars collide directly, leading to stronger shocks and a more energetic explosion that unbinds a larger fraction of the stellar material. 
Conversely, the grazing collision (Run 6) ejects less mass. Here, the interaction is confined to the low-density outer envelopes of the \texttt{MESA} stars, while the dense, tightly bound cores miss each other. 
This less disruptive encounter unbinds a smaller amount of material.

\begin{figure*}[htbp]
    \centering
    \begin{minipage}[b]{0.49\linewidth}
        \centering
        \includegraphics[width=0.95\textwidth]{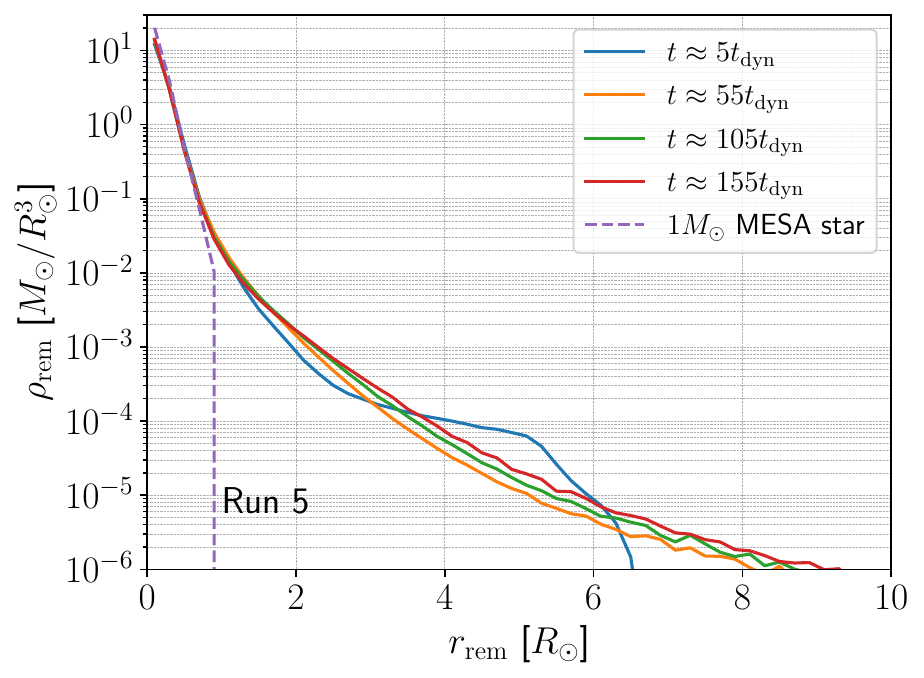}
    \end{minipage}
    \begin{minipage}[b]{0.49\linewidth}
        \centering
        \includegraphics[width=0.95\textwidth]{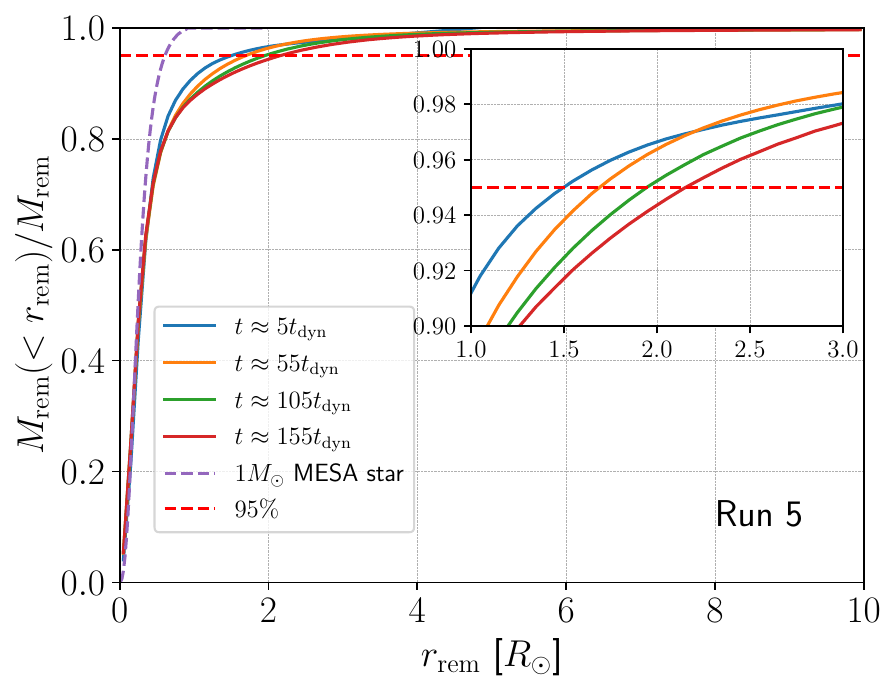}
    \end{minipage}
    \begin{minipage}[b]{0.49\linewidth}
        \centering
        \includegraphics[width=\textwidth]{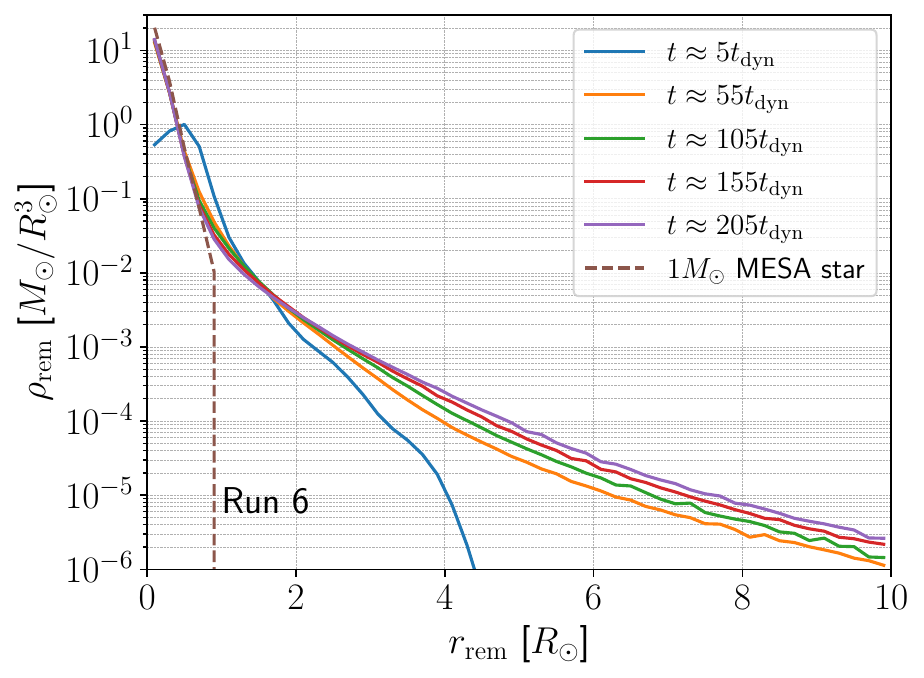}
    \end{minipage}
    \begin{minipage}[b]{0.49\linewidth}
        \centering
        \includegraphics[width=\textwidth]{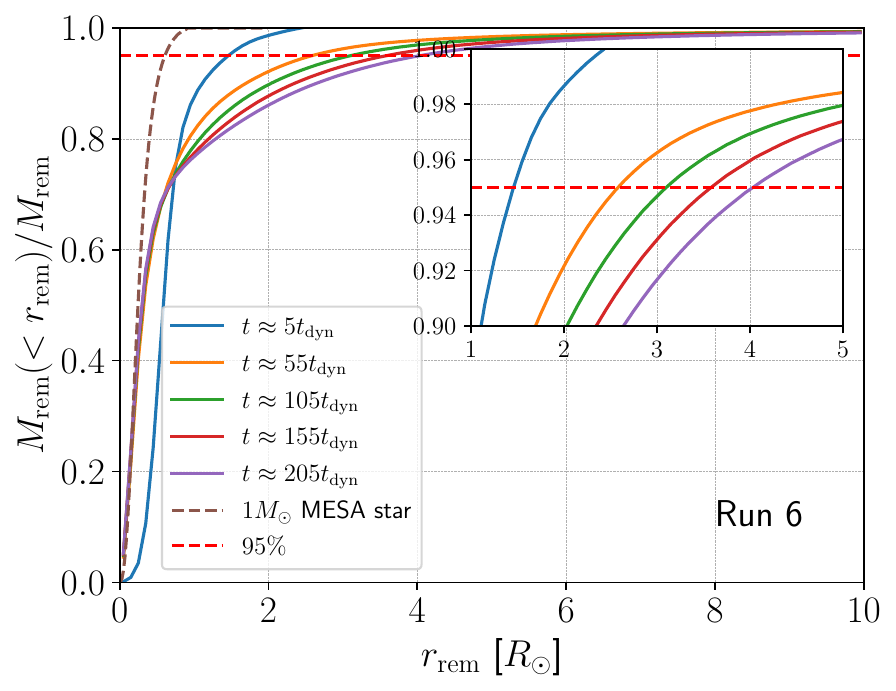}
    \end{minipage}
    \caption{Same as Figure~\ref{fig:headon_rho_m}, but for Runs 5 and 6. The dashed line represents the initial $1\,M_\odot$ MAMS profile obtained from \texttt{MESA}.
    The bump in the lower-left panel indicates the presence of the secondary star's core during the final inspiral phase, just before the full merger.}
    \label{fig:MESAgrazing_rho_m}
\end{figure*}

The internal structure of the final merger remnants is detailed in Figure~\ref{fig:MESAgrazing_rho_m}. 
A key feature, common to both head-on and grazing collisions, is that the remnant's central density is lower than that of the initial progenitor star, even though its total mass is nearly doubled. 
The immense kinetic energy dissipated during the collision causes the core to expand, resulting in a ``puffed-up" thermally supported remnant.
This core is surrounded by an extended envelope, qualitatively matching the structure seen in the polytropic mergers. 
Quantitatively, the remnant radii are slightly more compact than their polytropic counterparts, though they are still in a phase of slow, post-merger expansion driven by thermal relaxation.

\begin{figure*}[htbp]
    \centering
    \begin{minipage}[b]{0.49\linewidth}
        \centering
        \includegraphics[width=\textwidth]{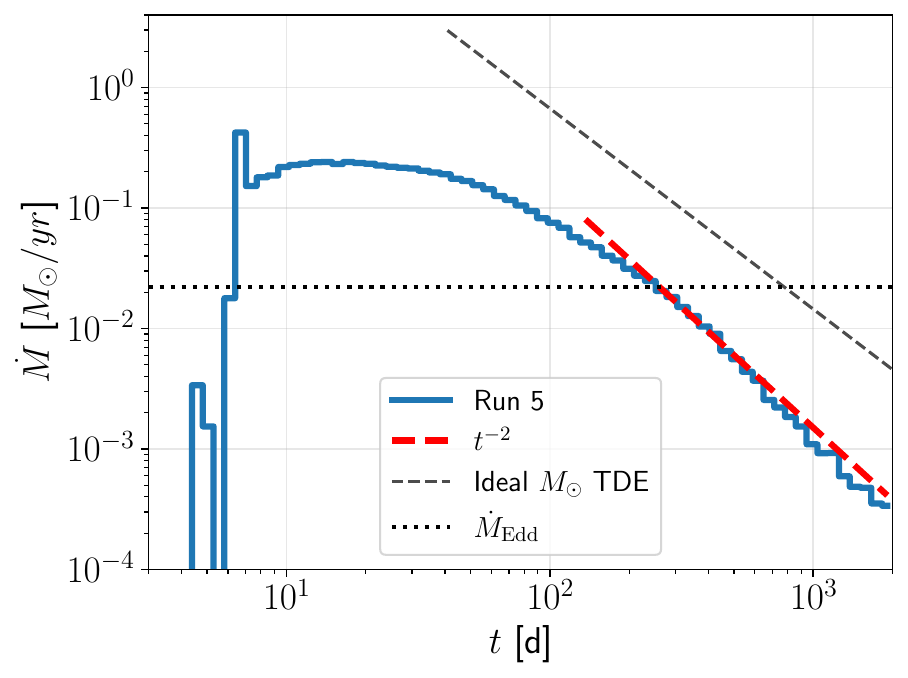}
    \end{minipage}
    \begin{minipage}[b]{0.49\linewidth}
        \centering
        \includegraphics[width=\textwidth]{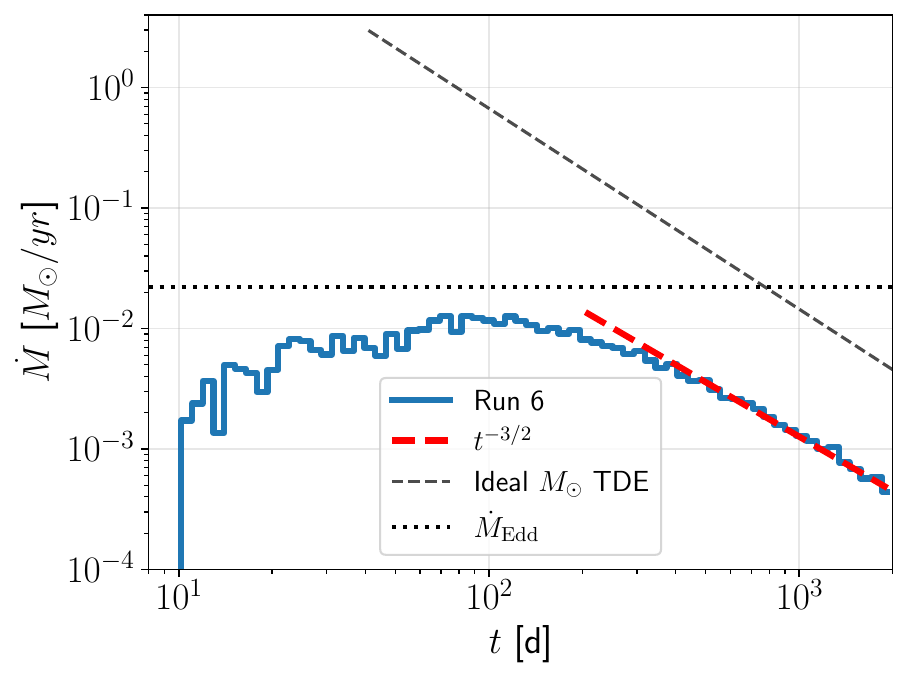}
    \end{minipage}
    \caption{Same as Figure~\ref{fig:fallbackrate}, but for Runs 5 and 6.}
    \label{fig:MESAfallbackrate}
\end{figure*}

Finally, Figure~\ref{fig:MESAfallbackrate} illustrates the mass fallback rates following the collisions. 
The magnitude of the fallback rate is correlated with the total mass loss. 
The head-on collision (Run 5), with its higher mass loss, exhibits a higher peak fallback rate. 
The grazing collision (Run 6) shows a lower rate. 
The temporal evolution also generally follows a power-law decay. 
A significant feature is the smooth fallback curve for the grazing merger (Run 6) - the grazing collision in this case leads to a single remnant. 
Hence, the fallback rate curve does not exhibit a dip as seen in the polytropic grazing case with two remnants (cf. Figure~\ref{fig:fallbackrate}, upper right panel).

\section{Summary and Discussion}
\label{sec:summary and discussion}

\subsection{Summary of Key Results}
\label{sec:summary}

In this paper, we have conducted a detailed study of the stellar collisions induced by the close encounter between a stellar binary and a SMBH. 
The binary, with a total mass $m_{12}$, an initial circular orbit, and a semi-major axis $a_b$, approaches the SMBH (mass $M_{\rm BH}$) on a parabolic orbit, with a pericenter distance $r_p$ from the SMBH.  
Using SPH simulations, we characterize key properties of stellar collisions and their potential observational consequences. 
We consider two type of stellar models: a $\gamma=5/3$ polytropic model (representing low-mass MS stars with mass $\lesssim 0.4M_\odot$) and a solar-like MS star obtained from \texttt{MESA}.

Our main findings focus on four representative cases involving both head-on and grazing collisions in gentle ($\beta_b \lesssim 1$) and deep ($\beta_b \gtrsim 1$) encounters (Table~\ref{tab:simulations}). 
We analyze the mass loss fraction, the density structure of the merger remnant(s), and the mass fallback rate.

\begin{enumerate}
    \item {\it Mass loss and remnant(s).}    
    Since the collision velocity is comparable to the stellar escape velocity (see Table~\ref{tab:simulations} and Figure~9 in Paper I), some mass loss is expected during the collision.
    Head-on collisions generally produce a single merger remnant, with a few percent mass loss (Figs.~\ref{fig:masslossfrac}, \ref{fig:MESAmasslossfrac}) 
    In contrast, grazing collisions exhibit more diverse outcomes. 
    For gentle encounters, while the first collision may not immediately result in a single merger remnant, the binary remains gravitationally bound; 
    the perturbed stars undergo multiple subsequent collisions (Figure~\ref{fig:gentle_grazing_sep}), ultimately forming a single remnant with minimal mass loss ($\lesssim 1\%$). 
    However, in deep encounters, where the binary system is already disrupted, if the first collision fails to produce a merger, two remnants can form, as seen in our Run 2, and the mass loss fraction is relatively higher ($ \sim 4\%$).
    Note that the outcome of grazing collisions in deep encounters is highly sensitive to the collision parameters and stellar structure. 
    While polytropic models result in two separate remnants, simulations with solar-type \texttt{MESA} stars yield a lower relative velocity and a tidal capture, resulting in a single merger remnant. 
    This suggests that for solar-type stars, mergers are a more common outcome than for low-mass MS stars.

    \item {\it Density structure of merger remnant(s).}    
    The remnant star formed after collision has a significant different structure compared to the original MS star. 
     The merger remnant is characterized by a highly extended envelope. 
    Its core structure depends on the progenitor model: for polytropic stars, the remnant has a much higher central density, whereas for $1 M_\odot$ \texttt{MESA} stars, the dissipated kinetic energy causes the core to expand, resulting in a lower central density than the initial star (Figs.~\ref{fig:headon_rho_m}, \ref{fig:grazing_rho_m}, \ref{fig:MESAgrazing_rho_m}).
    The extended envelope makes the merger remnant more susceptible to partial tidal disruption when it returns to the pericenter around the SMBH.
    If the two stars do not merge but instead undergo significant dynamical perturbations, the outcomes differ substantially (upper row of Figure~\ref{fig:grazing_rho_m}).
    These perturbed stars not only develop an extended envelope but also experience a substantial reduction in the central density. 
    Analogous to the Hills mechanism, one of the stars often remains bound to the SMBH. 
    In this case, this perturbed star is also significantly more susceptible to tidal disruption.

    \item {\it Orbital property of mass-loss debris and mass fallback rate.} 
    Nearly half of the mass-loss debris produced during the collision is expected to fall onto the SMBH, undergoing a process analogous to a TDE. 
    From the simulation snapshots (Figs.~\ref{fig:deepheadonmovie}-\ref{fig:gentlegrazingmovie}), we observe that the ``debris cloud" generated by the collision exhibits a morphology distinctly different from the ``debris stream" typical of TDEs. 
    Furthermore, an analysis of the binding energy distribution of the debris (Figs.~\ref{fig:massloss_energy},~\ref{fig:MESAmasslossfrac}) reveals significant deviations from the top-hat distribution observed in stellar TDEs.
    Specifically, the energy distribution is bimodal for collisions that produce a single remnant, and becomes triple-peaked for the cases that result in two separate remnants, with the dips corresponding to debris gravitationally bound to the remnant(s).
    Since the collision does not necessarily occur at the pericenter, unlike in the standard TDE, the resulting mass fallback accretion rate also differs (Figs.~\ref{fig:fallbackrate},~\ref{fig:MESAfallbackrate}).
    Although the total mass loss represents only a small fraction of the stellar mass and thus results in a relatively low magnitude of fallback accretion, the shape of the fallback rate curve is notably distinct from that of TDEs.  
     Specifically, the fallback curve shows notable features depending on the collision's outcome. 
    Generally, it remains approximately constant for an extended period ($\gtrsim 100$ days) before starting to decline (for $M_{\rm BH}=10^6M_\odot$ and solar-type stars). 
    For the two-remnant case arising from a deep grazing collision of polytropes, the fallback rate exhibits a distinct dip, caused by the bound remnant returning to its pericenter and re-capturing some of the debris (Figure~\ref{fig:fallbackrate}). 
    In contrast, for the \texttt{MESA} case in this work, which our grazing-encounter yields a single remnant, the fallback curve is smooth and lacks such a feature (Figure~\ref{fig:MESAfallbackrate}).
    Additionally, the angular momentum distribution of the mass-loss debris (represented by the debris' pericenter distance $r_p$, see Figure~\ref{fig:rp}) suggests that the resulting accretion disk extends to the original pericenter distance.

\end{enumerate}

\subsection{Parameter Space and Comments on the Results of Paper I}
\label{sec:discussion}

\begin{figure*}[htbp]
\centering
\begin{minipage}[b]{0.49\linewidth}
    \centering
    \includegraphics[width=\textwidth]{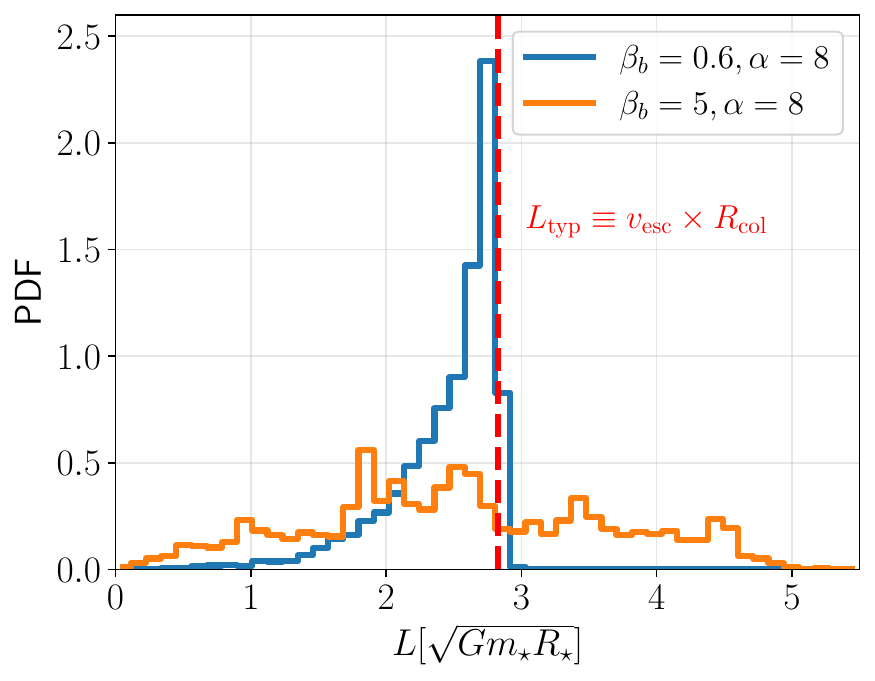}
\end{minipage}
\begin{minipage}[b]{0.48\linewidth}
    \centering
    \includegraphics[width=\textwidth]{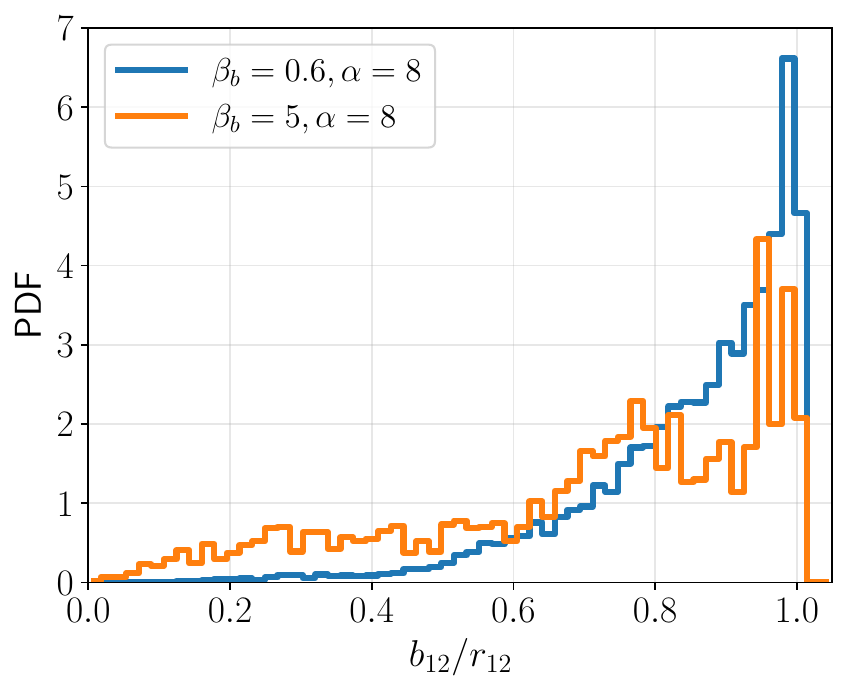}
\end{minipage}
\caption{
Probability density function (PDF) of the relative orbital angular momentum (left panel) and impact parameter $b_{12}/r_{12}$ (right panel) when two stars contact each other.
The blue and orange solid lines represent the results for two different values of $\beta_b$: gentle encounter ($\beta_b=0.6$) and deep encounter ($\beta_b=5$).
The red dashed line in the left panel represents the typical relative angular momentum for grazing collisions, where the contact velocity is the escape velocity $v_{\rm esc}=\sqrt{2Gm_{12}/R_{\rm col}}$ and impact parameter is $b_{12}=R_{\rm col}$.
}
\label{Fig:spin_L}
\end{figure*}

In this paper, we have carried out hydrodynamical simulations in four stellar collision cases (see Table~\ref{tab:simulations}), involving both gentle ($\beta_b=0.6$) and deep ($\beta_b=5$) encounters, leading to nearly head-on ($b_{12}/r_{12}\lesssim 0.06$) and more grazing ($b_{12}/r_{12}\simeq 0.5-0.9$) collisions.  
How representative are these cases? 
In Paper I, we considered a wide range of parameter space using N-body integrations, and calculated the probabilities of stellar collisions in binary-SMBH encounters; we adopted the inelastic collision assumption, which implies that all collisions lead to merger with no mass loss. 
Obviously, the collision probabilities and the impact velocity distribution obtained in Paper I remain valid.
The outcome of a collision, on the other hand, depends strongly on the collisional impact parameter ($b_{12}/r_{12}$).

To determine the range of outcomes of stellar collisions in binary-SMBH encounters, we show in Figure~\ref{Fig:spin_L} the probability distributions of the relative orbital angular momentum $L$ (left panel) and impact parameter $b_{12}/r_{12}$ (right panel) of the two colliding stars for both gentle and deep encounters, based on the simulation data of Paper I.
We see that most collisions are grazing collisions, characterized by relatively large $b_{12}/r_{12}$.  
The hydrodynamical simulations reported in this paper show that for such grazing collisions, the outcomes differ between the encounter types:
for gentle encounters, multiple collisions eventually lead to a single merger remnant; 
for deep encounters, the outcome is highly sensitive to the collision parameters and initial stellar structure. 
In our representative deep-grazing encounters, the polytropic case (Run 2) produces two perturbed remnants, whereas the \texttt{MESA} case (Run 6) produces a single merger. Because the contact conditions are not identical—see Table~\ref{tab:simulations}—we refrain from attributing the different outcomes to stellar structure alone. At contact, Run 2 has $v_{12}\approx1.354v_{\rm esc}$ and $b_{12}/r_{12}\approx0.489$, whereas Run 6 has $v_{12}\approx1.012v_{\rm esc}$ and $b_{12}/r_{12}\approx0.525$. Hence the contact kinetic energy in Run 2 is $\sim (1.354/1.012)^2\approx1.8\times$ that of Run 6, and the product $v_{12}(b_{12}/r_{12})$ is $\sim 25\%$ larger—both trends disfavoring capture/merger in Run 2. Small differences in contact speed and geometry can shift the balance among orbital binding, tidal energy deposition, and mass stripping; our results therefore illustrate the possible outcomes under these specific initial conditions, and a controlled survey varying one factor at a time will be needed to isolate the role of stellar structure. This interpretation is consistent with the classical tidal-capture scalings in which the dissipated energy rises steeply with decreasing pericenter distance \citep[e.g.,][]{PressTeukolsky1977,LeeOstriker1986}.

Nevertheless, the two key outcomes that are potentially observable -- mass loss and fallback accretion, and the formation of puffier stars -- remain largely unaffected. 
Even in cases with two perturbed stars, a fraction of the mass loss can still be accreted onto the SMBH, producing a flare, and the stars involved are significantly puffier than their initial states.  
Moreover, in such cases, one of the perturbed stars remains bound to the SMBH, typically on a shorter orbital period compared to the single merger remnant (see the comparison between Sections 6.2.1 and 6.2.2 in Paper I). 
This further supports the idea that stellar collisions induced by binary-SMBH encounters can enhance the likelihood of repeated partial tidal disruption events.

\vspace{5mm}
 We thank the anonymous referee for suggestions that have improved the manuscript.
This work made use of the Gravity Supercomputer at the Department of Astronomy, Shanghai Jiao Tong University.
The computations were supported by the National Science Foundation of China (grant No. 12173024).
\vspace{5mm}

\software{
\textsc{Phantom} \citep{Phantom},
\texttt{MESA} \citep{Paxton2011,Paxton2013,Paxton2015,Paxton2018,Paxton2019,Jermyn2023},
Sarracen \citep{Sarracen},
Matplotlib \citep{Hunter2007}, 
NumPy \citep{Walt2011}.
}

\bibliography{sample631}{}
\bibliographystyle{aasjournal}

\end{document}